# Clever algorithms for glasses work by time reparametrization

Federico Ghimenti,[1,2] Ludovic Berthier,[3,4] Jorge Kurchan,[5] and Frédéric van Wijland[1]

[1] *Laboratoire Matière et Systèmes Complexes (MSC), Université Paris Cité & CNRS (UMR 7057), 75013 Paris, France*

[2] *Department of Applied Physics, Stanford University, Stanford, CA 94305, USA*

[3] *Laboratoire Charles Coulomb (L2C), Université de Montpellier & CNRS (UMR 5221), 34095 Montpellier, France*

[4] *Gulliver, UMR CNRS 7083, ESPCI Paris, PSL Research University, 75005 Paris, France*

[5] *Laboratoire de physique LPENS, 24 rue Lhomond, 75005 Paris, France*

The ultraslow dynamics of glass-formers has been explained by two views often considered as mutually exclusive: one invokes locally hindered mobility, the other rests on the complexity of the configuration space. Here we show that time evolution responds strongly to details of the dynamics by changing the speed of time-flow: it has time-reparametrization softness. This finding reconciles both views: while local constraints reparametrize the flow of time, the global landscape determines relationships between different correlations at the same times. We show that modern algorithms developed to accelerate the relaxation to equilibrium act by changing the time reparametrization. Their success thus relies on their ability to exploit reparametrization softness. We conjecture that these results extend beyond the realm of glasses to the optimization of more general constraint satisfaction problems and to broader classes of algorithms.

In contrast to many fields of physics, acceptable microscopic models of glasses are easy to construct: hard spheres are a good example, and may be readily simulated or studied experimentally [1]. Theoretical efforts have instead been mostly directed at finding the right questions to ask [2]. Phenomenological arguments abound and are constantly being refined, while exact analytical results are rare, a recent notable exception being the case of particles living in an infinite-dimensional space [3, 4].

Several ideas have been proposed to capture the origin of the dramatic slowdown of the dynamics as density is increased or as temperature is lowered. These have been classified into two families. On the one hand, there is a landscape paradigm, where the energy function and associated static expectation values encode for the dynamical behavior [5–10]. A dynamic transition, where the relaxation time diverges at a finite temperature, is then only possible if there exists an underlying, thermodynamic and landscape-based, one [11–16]. On the other hand stands the view that glassiness is dynamical in essence. An example is when local rearrangements, and hence flow, are dominated by dynamic facilitation [17, 18]. This occurs when localized, mobile regions diffusing through the system are essentially non-interacting and cannot be born or die except by branching and coalescing, a mechanism absent from the high-dimensional solution [19, 20].

The latter view and the statics-dynamics debate were recently boosted by the development of numerical algorithms that drastically accelerate the dynamical evolution [21–23], while preserving the Boltzmann distribution. In particular, the outstanding performance of the swap Monte Carlo algorithm has been interpreted as indicating that the energy landscape paradigm is insufficient to account for glassiness and thus as a sign that real glasses behave as models with local kinetic constraints [19, 24, 25].

Our central finding is that both pictures are simultaneously right. That there is no contradiction between them is the consequence of a remarkable property, time reparametrization invariance, that is asymptotically exact – i.e. if and when the dynamics are slow – in the mean-field theory of aging in disordered systems [26–29]. For realistic, finite-dimensional models, we are at present far from being able to prove analytically that time reparametrization invariance holds. Instead, we may seek whether the associated time reparametrization softness [30–34] is observed in them. If true, this implies that the flow of time can be modified with weak perturbations, exactly like a film being projected using an old hand-cranked projector, where the operator can effortlessly modify the speed of the projection but has no control over the film's content. Physically, this implies that local particle rearrangements and short-time relaxation can be different depending on the specific dynamics and algorithm employed. Each algorithm yields its own collective relaxation time, but they trace the same trajectory in configuration space as the system slowly and collectively relaxes. This description echoes earlier attempts to remove time from the description of relaxation phenomena [35–38]. Overall, our results provide a clear means of sieving out the energy-landscape-dependent features from purely dynamical ones.

Our presentation begins with numerical and analytical observations of parametric plots of correlations in glass-formers endowed with vastly different time evolutions but sharing the same stationary Boltzmann distribution. The spectacular impact of reparametrizations calls for theoretical support that we address with an exact solution in mean field and eventually by describing a general picture with consequences well beyond the physics of glasses.

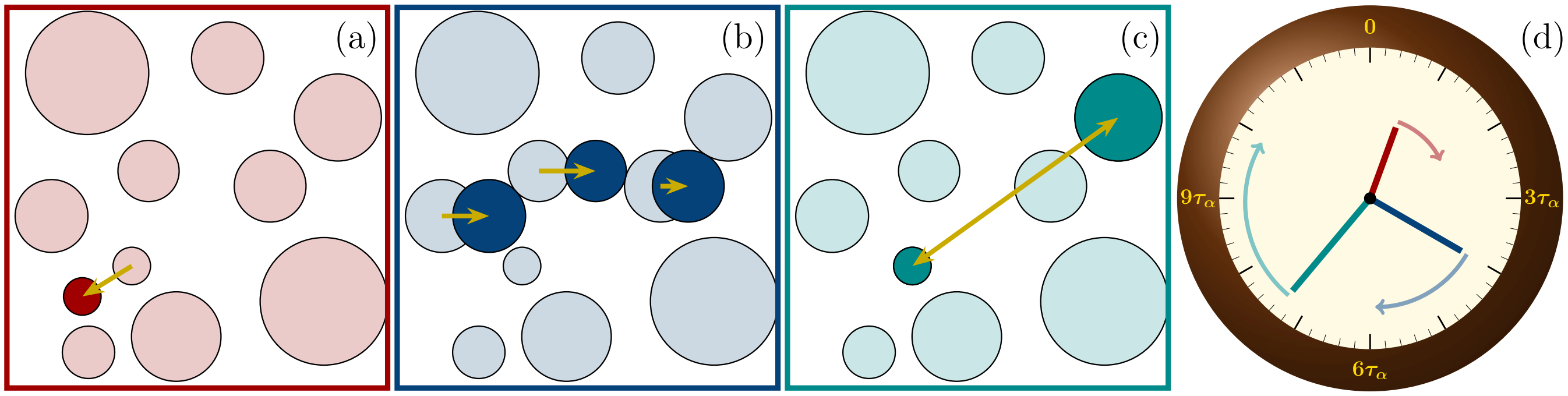

FIG. 1. Sketch of three algorithms performing dramatically different moves on the same system: (a) local equilibrium translations; (b) directed displacement of chains of particles; (c) exchange of particle diameters. The effect of these moves can be encapsulated into a change of the pace at which the clock of the system is ticking. This is time reparametrization softness.

## THE PROOF IS IN THE PUDDING

### Parametric plot of time correlations

As glassiness sets in, the decay of time correlations follows a two-step scenario beginning with a fast relaxation, followed by a slow decay that extends up to a characteristic time $\tau_\alpha$. We track the relaxation of the system in terms of a real space, collective overlap function $F_o$, defined by

$$F_o(a,t) \equiv \frac{1}{N}\sum_{i,j}\langle\theta\left(a - |\mathbf{r}_i(t) - \mathbf{r}_j(0) - \Delta\mathbf{r}_{\rm com}(t)|\right)\rangle, \quad (1)$$

where $\Delta\mathbf{r}_{\rm com}(t)$ is the displacement of the center of mass of the system, and $\theta$ is the step function. Physically, the function $F_o$ compares the density fields of the system between times 0 and $t$. The angular brackets represent an average over the Boltzmann initial state and over time realizations of the dynamics. The parameter $a$, chosen to be smaller than any particle's diameter, selects the scale over which the relaxation is monitored. This is better suited than the usual self-intermediate scattering function (see SI [39]) for discussing various algorithms on equal footing. In practice, we plot a connected and normalized overlap function $Q_o(a,t) = (F_o - q_r)/(1-q_r)$, with $q_r = \frac{4\pi}{3}\rho_0 a^3$ in three dimensions, and $\rho_0$ the density of the system. This function decays to 0 at large times for any value of $a$ and is normalized to unity at $t = 0$. The statements we make concern the late stages of the relaxation process near $\tau_\alpha$.

When the goal is to sample the Boltzmann distribution, the algorithm being used does not need to respect actual physical constraints. For instance, introducing particle swaps with detailed balance, while clearly unphysical, has been shown to drastically reduce $\tau_\alpha$. Other proposals rely on irreversible methods that affect translation moves and/or particle swaps. The overall two-step relaxation shape of $Q_o(a,t)$ is also found in these vastly different dynamical evolutions. Our central result, sketched in Fig. 1, is that all dynamics, fast and slow, encapsulate the same physical evolution of the correlations. To test time reparametrization softness, we eliminate time in parametric plots of $Q_o(a,t)$ for two values of $a$ and for various dynamics in different systems with glassy behavior. The collapse of the curves then directly expresses time-reparametrization invariance. Our conjecture is that this collapse occurs whenever both the algorithms display slow relaxation, with long lived plateaus of the same height in the decay of their dynamical correlation functions. The numerical evidence supporting these claims is our main result.

### Irreversible translation moves

We begin with a three-dimensional binary Kob-Andersen mixture [40] evolving through an overdamped Langevin process. The convergence of this dynamics to the same Boltzmann distribution can be accelerated by exerting on every particle an extra force transverse to its local energy gradient [41–43]. We use the dimensionless parameter $\gamma$ to quantify the relative strength of transverse to radial forces: the larger $\gamma$, the greater the speed-up (see Methods for further details).

We numerically integrate the dynamics at a low temperature $T = 0.48$, where the relaxation spans several orders of magnitude. In Fig. 2(a, b) we plot the time evolution of $Q_o$, evaluated at two distinct lengths $a_1$, $a_2$. In the presence of transverse forces, the relaxation of the system is faster compared with the equilibrium dynamics by a factor $\approx 4$ for $\gamma = 4$ and by a factor $\approx 7$ for $\gamma = 10$. Both dynamics exhibit signs of two-step relaxation, and in the late part of the relaxation the shape of the curves is very similar, even when passing from $\gamma = 0$ to $\gamma = 10$. In Fig. 2(c), we represent $Q_o(a_1,t)$ as a function of $Q_o(a_2,t)$ in a parametric plot, for $\gamma = 0, 4$ and 10. Their collapse shows that there exists a reparametrization of time such that the slow relaxation of the system with transverse forces coincides with the slow relaxation of the system under equilibrium dynamics: transverse forces speed up the slow dynamics of the system by means of a time reparametrization. Recent results [41–43] demonstrated that at low temperatures the dynamic pathways involve genuine non-equilibrium currents. Yet, these dynamic pathways display time reparametrization.

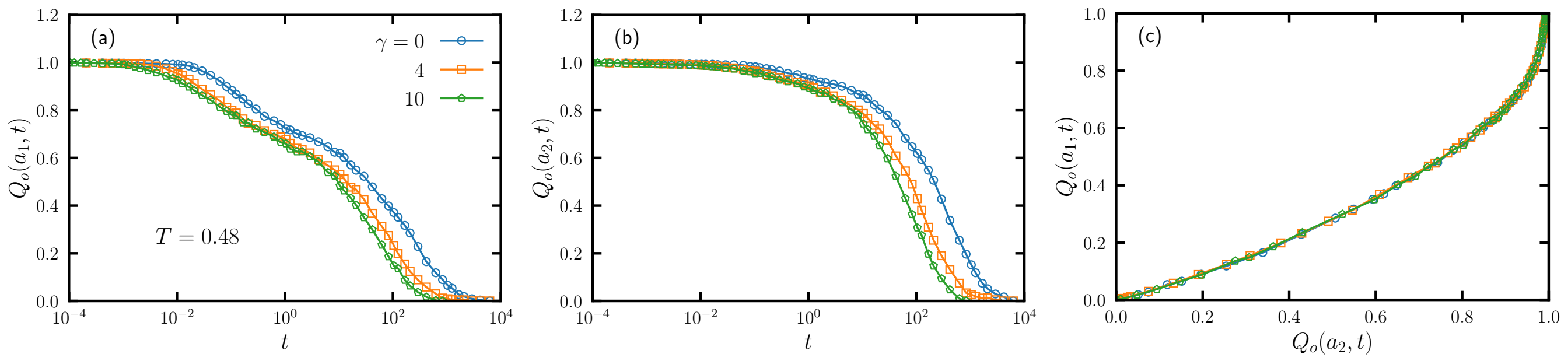

FIG. 2. (a-b) Time evolution of the overlap function $Q_o(a,t)$ for a Kob-Andersen mixture under equilibrium overdamped Langevin dynamics ($\gamma = 0$) and with transverse forces ($\gamma = 4$ and $\gamma = 10$), for $a_1 = 0.2d_{AA}$ and $a_2 = 0.3d_{AA}$. (c) Time reparametrization invariant plot, obtained representing $Q_o(a_1,t)$ as a function of $Q_o(a_2,t)$.

We have also compared the dynamics of polydisperse hard spheres with the equilibrium Metropolis Monte Carlo algorithm against the irreversible Event Chain Monte Carlo [44]: a protocol of driven and collective rejection-free displacements of chains of particles which are microscopically very different from the local moves of the Metropolis Monte Carlo algorithm. We discovered that Event Chain also operates in glassy dynamics by reparametrizing the time over which the long-time relaxation occurs. The numerical data supporting this statement are shown in the SI [39].

### Swap Monte Carlo algorithm

We push our exploration of time reparametrization invariance by probing the Swap Monte Carlo algorithm [21] (hereafter denoted as "Swap"). Swap implements reversible exchanges of diameters between pairs of particles. For continuously polydisperse hard-spheres, this allows the particle diameters to fluctuate, potentially opening up additional relaxation channels [20, 45–47] that can speed up the dynamics by several orders of magnitude [22, 48]. However, we claim that in the region where the dynamics is slow for both Metropolis and Swap algorithms, the acceleration achieved by Swap results again from a reparametrization of time with respect to the local Metropolis dynamics.

To see this, we compare Metropolis and Swap dynamics in a dense system of continuously polydisperse hard spheres. We tune the rate of swaps $p_s$ to work in a regime where the speedup provided by Swap encompasses about four orders of magnitude, and track the relaxation by means of the overlap function $Q_o(a,t)$. In Figs. 3(a-c), we show the time evolution of $Q_o(a,t)$ for three different values of $a$. For Metropolis dynamics, the decay of $Q_o$ covers several orders of magnitude, with signs of a two-step relaxation showing up for the lowest value of $a$. Turning on the swap moves, the curves depart from the ones obtained with the Metropolis dynamics, and their decay is faster.

In Figs. 3(d-f), we show parametric plots of $Q_o(a_i,t)$ as a function of $Q_o(a_j,t)$ for all possible combinations $(a_i,a_j)$ of distinct parameters for the overlap function. All the curves for the dynamics with different swap probabilities nearly collapse on top of each other, and the four-order of magnitude speed up disappears, suggesting that even an algorithm as powerful as Swap works by reparametrizing time. We have observed that the quality of the collapse improves when density is increased and all curves display enhanced sign of two-step decay, suggesting even better collapse could be observed if we could simulate longer timescales. We expect that the time window over which data collapse is observed increases when the average relaxation becomes larger, an observation that would deserve further work.

Irreversible versions of Swap have recently been proposed [49, 50] that further accelerate the dynamics by performing driven, collective exchanges of particle diameters. The results are as above, see SI [39] for data supporting our findings.

### Kinetically constrained model

We now provide an instance of glassy dynamics that does not display time reparametrization softness because it is controlled by purely local constraints. We study the dynamics of the soft-East kinetically constrained model introducing an analog of swap moves, as proposed in Ref. [24]. Without softness, the original East model consists of $N$ binary variables $n_i \in \{0,1\}$ on a one-dimensional periodic lattice. Sites with $n_i = 1$ are occupied by an excitation. In the spirit of local defects in crystal, excitations represents regions where structural rearrangment can take place, transitioning from $n_i = 1$ to $n_i = 0$. The thermodynamic properties are trivial, and are governed by the temperature $T$ and the energy cost $J$ of an excitation. The relaxation is hindered by kinetic constraints: the reversible creation and destruction of an excitation can take place only on a site immediately to the right of an excitation, thus incorporating dynamical facilitation.

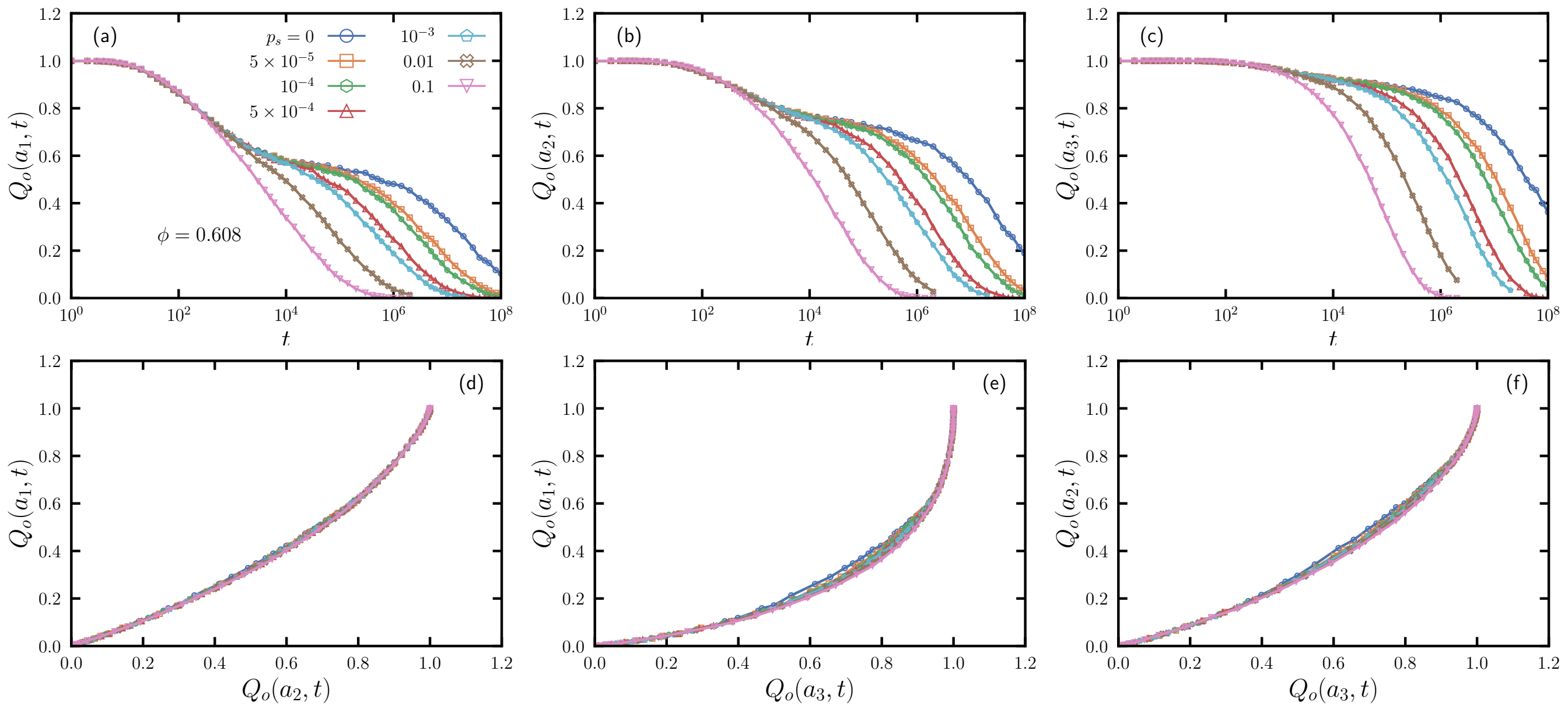

FIG. 3. (a-c) Time evolution of the collective overlap function in continuously polydisperse hard spheres using Metropolis algorithm ($p_s = 0$) and Swap Monte Carlo with different swap probabilities at fixed packing fraction $\phi \equiv \rho_0 \frac{\pi}{6} \overline{d^3}$. In each panel, a different value of $a$ is used. $a_1 = 0.15\overline{d}$, $a_2 = 0.2\overline{d}$, $a_3 = 0.3\overline{d}$. (d-f) Parametric plots of $Q_o(a_i, t)$ as a function of $Q_o(a_j, t)$.

In the soft version, each site is supplemented with a binary softness parameter, which modifies the kinetic constraint, allowing for the birth and death of isolated excitations. Spontaneous updates of the softness are controlled by a swap-like process with rate $r_s$. When $r_s = 0$, only the softness of already existing excitations can change. For $r_s \neq 0$, the softness of any site can be updated (see Methods for more details). By changing $r_s$ we can thus speed the dynamics up and investigate whether time reparametrization is at work in the system.

We simulate the dynamics of the soft-East model with swap updates. For a fixed temperature we change the softness update rate from $r_s = 0$ to large values. We track the relaxation of the system using the persistence function $P(t)$, which measures the fraction of spins that have not yet flipped up to time $t$, and is thus analogous to the overlap function used for structural glasses. The evolution of $P(t)$ with $r_s$ is shown in Fig. 4(a). The persistence function starts from 1 when no spin has flipped and decays to 0 when all spins have been updated at least once. The behavior of $P(t)$ at short times does not change upon varying $r_s$. At later times, the curves for $r_s \neq 0$ depart from the curve obtained for $r_s = 0$, achieving a speedup of almost three orders of magnitude for the largest $r_s$. We also looked at the time autocorrelation of the spins, $C_n(t)$, see Fig. 4(b). For any swap rate $r_s$, the decay of $C_n(t)$ is about an order of magnitude faster than the decay of the corresponding persistence curve $P(t)$, but the overall behavior is very similar.

The long time decay of $P(t)$ and $C_n(t)$ have different shapes when $r_s$ changes. In Fig. 4(c) we show parametric plots of $C_n(t)$ against $P(t)$. The different curves do not collapse, even for the lowest $r_s$ value. Differently from the particle models, the correlation functions studied in Fig. 4 are single spin functions, as collective correlations vanish for this model. To assess whether the access to nonvanishing collective correlations on different length-scales can restore time-reparametrization invariance, we investigate in the SI a different kinetically constrained model with softness updates [39] where collective correlations can be studied. We do not see a collapse in the parametric plots in that system either. The acceleration provided by the additional updates of the softness are not simply described by reparametrization of time. This also shows that the reparametrization-invariant data collapse observed in finite dimensional particle models is not trivial.

### Time reparametrization softness

Mean-field glass models provide a tractable case (or limit, in the case of large dimensionality) where there is analytic support to the concept of time reparametrization. We first describe how time reparametrization softness generally arises in mean-field frameworks, and then tackle analytically a specific mean-field model.

Glasses primarily respond to external perturbations by changing the pace of their evolution, like the same movie projected at various speeds. Consider a correlation function obtained by measuring a quantity at two times and repeating the experiment many times under statistically equal conditions, $C_A(t,t') = \langle A(t)A(t')\rangle$. (Averages are over independent experiments). Alternatively, we may apply an infinitesimal pulse field changing the Hamiltonian to $H(t) \to H + hA\delta(t-t')$ at time $t'$ and measure the change in the expectation at time $t$ at linear order, $R_A(t,t') = \frac{\delta\langle A(t)\rangle}{\delta h}$. In equilibrium, the response and correlation functions are linearly related by the fluctuation-dissipation theorem, $TR_A(t-t') = -\partial_t C_A(t-t')$, with $T$ the temperature of the system.

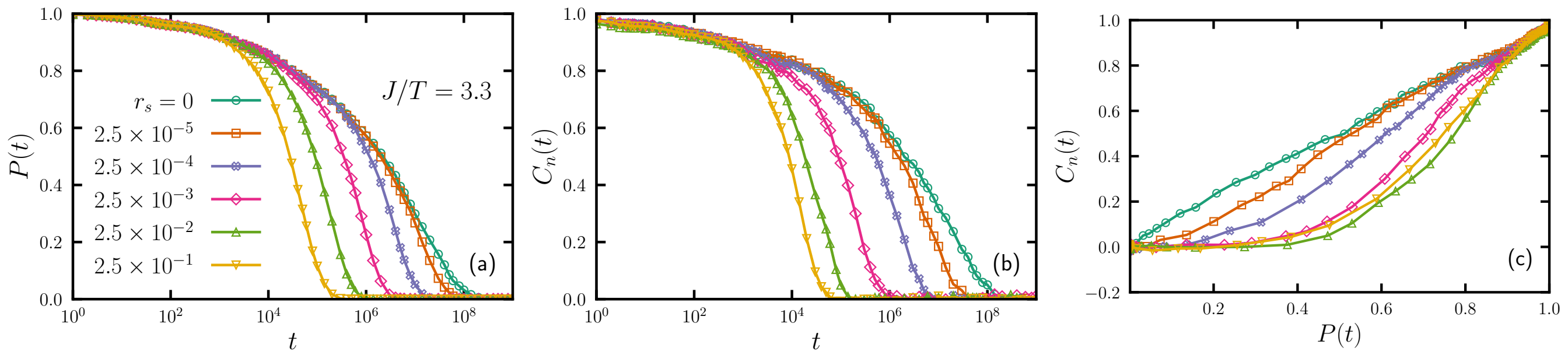

FIG. 4. Dynamics of a soft-East kinetically constrained model with softness updates, using different values of the softness updated rate $r_s$. (a) persistence function as a function of time for different swap rates. (b) Time decay of the spin spin autocorrelation function $C_n$. (c) Parametric plot of $C_n$ against $P$. The absence of collapse in (c) reveals the lack ot time reparametrization invariance.

In mean-field models, given a set of observables $A_\alpha$, it is possible to write closed exact equations for pairs of response-correlation functions $C_{A_\alpha}$ and $R_{A_\alpha}$, valid in and out of equilibrium, which do not have any explicit time dependence. If the relaxation of the system is very slow, one may neglect the time-derivatives and solve for all the $C_{A_\alpha}$, $R_{A_\alpha}$. The solution holds however *up to a time reparametrization*: if $\{R_{A_\alpha}(t,t'), C_{A_\alpha}(t,t')\}$ is a solution, then $\{R_{A_\alpha}(h(t),h(t'))\frac{\mathrm{d}h}{\mathrm{d}t'}, C_{A_\alpha}(h(t),h(t'))\}$ is also a solution. The function $h(t)$ is smooth and increasing, and it encodes the adopted reparametrization of time.

The true solution, where the time derivative is not neglected, is unique, and hence only one of the $h(t)$ constitutes the 'good' parametrization. However, other reparametrizations can be selected, for instance in the presence of an applied shear [51], or during jump events [34] between distant configurations at low energy, and can be made visible when looking at fluctuations of the correlation-response curves [27–29].

By making a parametric plot of any one of the $\{R_{A_\alpha}(t,t'), C_{A_\alpha}(t,t')\}$ in terms of a single, reference correlation $C_o(t,t')$, we obtain a single master curve independent of the chosen time reparametrization [52]. In this representation, we are measuring everything in terms of a 'clock' $C_o(t,t')$, which determines the 'material time' [35, 36, 38, 53].

The concept of time reparametrization first arose in the context of dynamic mean-field theory of aging systems [54]. As discussed here and in the specific mean-field model solved below, it generalizes to mean-field models in a wide range of situations, and it appears as a generic property of mean-field glasses in a slow dynamical regime. The numerical results provided in this work demonstrate that its domain of applicability is not restricted to mean-field models, but is an observed property of realistic, finite dimensional glass models. This is reminiscent of the idea of an effective temperature [55] that first emerged from the analytic solution of the aging dynamics of a mean-field spin glass model [54]. Along the years, the existence of effective temperatures was then demonstrated in a much wider range of finite dimensional situations [56].

### A solvable model

The above general discussion can be illustrated on an explicit example. We consider the mean-field $p$-spin glass model, which is driven out of equilibrium while respecting the Boltzmann distribution. To do so, we couple two copies of the system by means of a nonreciprocal force chosen to preserve the factorized Boltzmann distributions for the two systems [57]. As for transverse forces, a dimensionless parameter $\gamma$ controls the relative strength of the antisymmetric coupling. While the non-reciprocal coupling accelerates the dynamics, we can establish that time reparametrization holds exactly and is entirely governed by the equilibrium free energy landscape.

Below a critical temperature $T_d$, the ergodicity of the system is broken. Here, we investigate the role of time reparametrization invariance by considering the dynamics of the coupled systems at a temperature slightly above or slightly below $T_d$, starting from an initial configuration at infinite temperature. The fully connected nature of the model allows us to study its full time evolution in terms of spin-responses and spin-spin correlations $R_{\alpha\beta}(t,t')$ and $C_{\alpha\beta}(t,t')$, for $\alpha,\beta = 1,2$. Our analysis (details in [39]) shows that the slow evolution of $R_{\alpha\beta}(t,t')$ and $C_{\alpha\beta}(t,t')$ is fully determined by two scalar quantities $\widetilde{C}(t,t')$, $\widetilde{R}(t,t')$, which satisfy integral equations of

the form:

$$\begin{aligned} \mathcal{F}[\widetilde{C}(t,t'),\widetilde{R}(t,t')] &= 0 \\ \mathcal{G}[\widetilde{C}(t,t'),\widetilde{R}(t,t')] &= 0, \end{aligned} \tag{2}$$

with $\mathcal{F}$ and $\mathcal{G}$ functionals of $\widetilde{C}(t,t'),\widetilde{R}(t,t')$ which do not depend on $\gamma$. They are the same functionals found using equilibrium dynamics at $\gamma = 0$ [51, 54]. (2) determines $\widetilde{R}(t,t')$ and $\widetilde{C}(t,t')$ up to a reparametrization of time [27, 51]. The choice of the specific solution of (2) is done by matching asymptotically the slow terms with the fast decaying part, whose evolution does depends on $\gamma$. The impact of $\gamma$ on the slow decay of the correlations and response functions thus amounts to a time reparametrization.

## THE BIG PICTURE: FRANZ PARISI POTENTIAL AND QUASI-DYNAMICS

The stimulating picture emerging from our results is that both the energy landscape and the specifics of the dynamics matter. The former determines the form of the relaxation, while the latter controls the speed at which the configuration space is explored. We now provide a broader physical interpretation of the presence of time-reparametrization invariance, inspired by mean-field results.

In a given system, we introduce a notion of correlation between two configurations $\boldsymbol{x}$ and $\mathbf{y}$, such as

$$F_o(a,\mathbf{x},\mathbf{y}) = \frac{1}{N}\sum_{i,j}\langle \theta\left(a - |\mathbf{x}_i - \mathbf{y}_j|\right)\rangle, \tag{3}$$

where $\boldsymbol{x}$ is drawn from the Boltzmann distribution. We further consider the Boltzmann distribution of $\mathbf{y}$ at fixed $\mathbf{x}$, restricted to the surface $F_o(a,\mathbf{x},\mathbf{y}) = q$, where $q$ measures correlations between the two configurations. The corresponding $q$-dependent free energy $V(q)$ is the Franz-Parisi potential [58]. A sketch of $V(q)$ for different temperatures is shown in Fig. 5(a-c). Within mean-field, the potential is monotonic in the liquid phase and develops, at the dynamic transition point, a secondary minimum, whose location defines the Edwards-Anderson parameter $q_{EA}$. This minimum decreases until it becomes degenerate with the one at $q = 0$ at the equilibrium transition to the glass phase.

In Fig. 5(d-f) we sketch what would happen if we fixed two distances. Clearly, integrating one constraint away gives back the original one-dimensional potential. The assumption behind this picture is that, apart from the minimum at the origin, there is at most only another minimum in this potential at the position $(q_{EA}^1, q_{EA}^2)$.

Starting from here, we can make a 'quasidynamic' construction step by step, as sketched in Fig. 5(g). Very surprisingly, this construction gives, within mean-field, the correct reparametrization-invariant equations for the dynamics [59, 60], just by interpreting the links in the chain as times. From this construction, the chain is independent of the choice of correlation used for the links. Thus, we have found a relation between correlations that is a result of a purely equilibrium calculation, as no allusion was made to an actual dynamical process. As a corollary, our numerical results in finite dimension support the existence of quasi-dynamics robust with respect to time evolution. For transverse forces and Event Chain Monte Carlo algorithm, recent analytical and numerical results substantiate the quasi-dynamics picture [41, 49]: in transverse forces, odd transport significantly enhances local motion on short time-scales, but the particles eventually need to confront cage escape. In Event Chain Monte Carlo, maps of dynamical hetereogeneities over long time scales have been shown to be very similar to the ones obtained with Metropolis Monte Carlo.

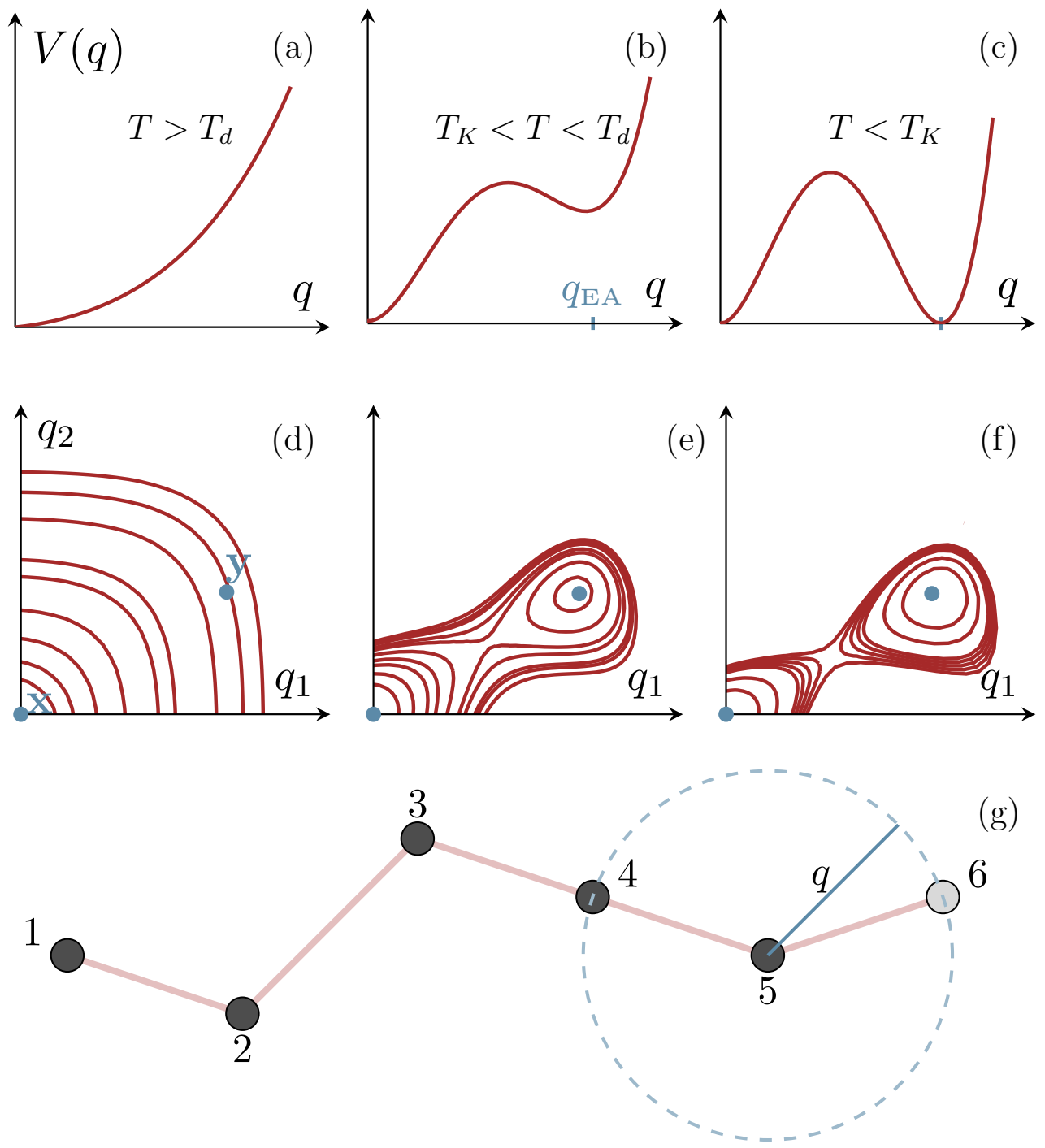

FIG. 5. Sketch of the Franz-Parisi construction in mean field: (a) above the dynamic transition temperature $T_d$. (b) Between the dynamic transition temperature and the static transition temperature $T_K$. (c) Below the static transition. (d-f) Sketch of the contour plots of a two dimensional Franz-Parisi potential, with two 'distances' $q_1$ and $q_2$ imposed. (g) Quasi-dynamics construction [59]: at each step we choose a new configuration, represented by the gray circles, subject to one or more constraints, keeping all the preceding links frozen.

What we have argued in this paper is that clever algorithms work by efficiently exploiting the softness implied by time reparametrization invariance to drastically accelerate the time evolution. Numerically probing, in finite-dimensional systems, the 'quasi-dynamics' picture of time-reparametrization put forth in our work is perhaps within reach, *e.g.* by building upon previous efforts along this direction [61–63].

Let us emphasize that the conclusion here is that the fact that different dynamical procedures leading to the same equilibrium may drastically stretch the timescales – the 'reparametrization softness' – is an observed property, one that any glass theory is required to reproduce. Even though it arises naturally in mean-field theory, it may perhaps be explained within other scenarios. It is an interesting question to understand, for instance, how this may arise, as it should, in a theory based on local elasticity.

More broadly, reparametrization softness has recently been identified as the mechanism leading to the emergence of gravity as a low-energy limit of simple quantum (SYK) models [33]. Hints of time reparametrization softness have been also found in the learning dynamics of wide and deep neural networks [64]. In the context of supercooled liquids, the very same mechanism underlies the drastic time-rescaling with temperature (as described by the time-temperature superposition principle), as well as under shear, aging, and barrier-crossing processes, with experimental consequences that are starting to be explored [35]. We hope that the present work incentivizes the use of parametric plots as a tool to probe time reparametrization in a variety of experimental settings. In this work, we demonstrate that reparametrization softness also resolves the longstanding dichotomy between dynamical and landscape views on glasses: the latter determines reparametrization-invariant characteristics, while the former governs the actual time parametrization.

## ACKNOWLEDGMENTS

We warmly acknowledge very useful discussions with Rob Jack, Juan P. Garrahan, Giorgio Parisi, and Pierfrancesco Urbani. LB, FG and FvW acknowledge the financial support of the ANR THEMA AAPG2020 grant. FG acknowledges support from Stanford's Leinweber Institute of Theoretical Physics.

## DATA AVAILABILITY STATEMENT

The data used to produce the figures in this paper are publicly available in a Github repository [65].

## MATERIALS AND METHODS

### Kob-Andersen mixture with transverse forces

The Kob Andersen potential is defined as [40]

$$V_{ab}(r) = 4\epsilon_{ab}\left[\left(\frac{d_{ab}}{r}\right)^{12} - \left(\frac{d_{ab}}{r}\right)^{6}\right] + C \tag{4}$$

for $r_{ab} < 2.5d_{ab}$, and 0 otherwise. The constant $C$ ensures that $V_{ab}(2.5d_{ab}) = 0$. The interaction diameters $d_{ab}$ are $d_{11} = 1$ (which sets the units of length of the system), $d_{12} = d_{21} = 0.8$, $d_{22} = 0.88$, while the interaction energies $\epsilon_{ab}$ are $\epsilon_{11} = 1$ (which sets the units of energy), $\epsilon_{12} = \epsilon_{21} = 1.5$, $\epsilon_{22} = 0.5$.

The overdamped Langevin dynamics for this system is [41]

$$\dot{\mathbf{r}}_i^a = -(\mathbf{1}+\gamma\mathbf{A})\sum_{j\neq i}\sum_{b=1}^{2}\boldsymbol{\nabla}_i^a V_{ab}(|\mathbf{r}_i^a-\mathbf{r}_j^b|)+\sqrt{2T}\boldsymbol{\xi}_i^a(t), \tag{5}$$

where $\mathbf{r}_i^a$ is the position of particle $i$ of species $a$, with $a = 1, 2$. $\boldsymbol{\xi}_i^a(t)$ is a Gaussian white noise with zero mean and correlations $\langle\boldsymbol{\xi}_i^a(t)\otimes\boldsymbol{\xi}_j^b(t)\rangle = \mathbf{1}\delta(t-t')\delta_{ij}\delta_{ab}$. The temperature $T$ is measured in units of $\epsilon_{11}/k_{\mathrm{B}}$, with $k_{\mathrm{B}}$ the Boltzmann constant. The transverse forces are implemented by means of an antisymmetric matrix $\mathbf{A} = \begin{bmatrix} 0 & -1 & 0 \\ 1 & 0 & 0 \\ 0 & 0 & 0 \end{bmatrix}$ and of a dimensionless number $\gamma$ that controls the amplitude of the nonequilibrium drive. When $\gamma \neq 0$, (5) is an out of equilibrium dynamics, with a steady state given by the Boltzmann distribution for the Kob-Andersen mixture. The relaxation to the stationary state for $\gamma \neq 0$ is guaranteed to be shorter than or equal to the one of equilibrium dynamics.

Simulations are performed in the NVT ensemble, using a box of side $L = 9.4d_{11}$ and a total number of particles $N_1 + N_2 = 1000$, so that the number density of the system $\rho_0 = \frac{N}{V}$ is $\rho_0 \simeq 1.204$. The equations of motion given by (5) are integrated by means of the Euler-Heun algorithm using a time step $\Delta t = 10^{-4}$. The data shown are obtained from the steady state dynamics equilibrated samples, obtained using the overdamped Langevin dynamics ( (5) with $\gamma = 4$) for $10^8$ time steps. The results shown in Fig. 2 of the main text are obtained by studying the stationary dynamics of 45 independent configurations.

### Polydisperse hard spheres

The model consists of $N$ polydisperse hard spheres in three dimensions [48]. The diameters $d$ are drawn from a power law distribution $\pi(d) \propto d^{-3}$. The boundaries of the distribution are chosen so that the polydispersity $\Delta \equiv \frac{\sqrt{\overline{d^2}-\overline{d}^2}}{\overline{d}}$ is $\Delta \approx 23\%$, with $\overline{\cdots}$ denoting an average over the diameter distribution. The average diameter $\overline{d}$ sets the units of length. The hard sphere potential between two particles $i$ and $j$ separated by a distance $r_{ij}$ is defined as $V(r_{ij}) = +\infty$ if $r_{ij} < \frac{d_i+d_j}{2}$, and $V(r_{ij}) = 0$ otherwise. The simulations are done in a cubic box of linear size $L$ with periodic boundary conditions. We explore the dynamics of the system at high packing fractions $\phi \equiv \frac{\pi}{6}\frac{N}{L^3}\overline{d^3}$.

### Metropolis algorithm

In a single timestep of the Metropolis algorithm, $N$ Metropolis moves are performed. During a Metropolis move, a sphere is selected uniformly at random, and a displacement is proposed within a cubic box of side $\delta$, centered around the sphere. If the displacement does not generate any overlap between the sphere and its neighbors, the move is accepted. In our simulations, we chose $\delta = 0.115$.

### Event-Chain Monte Carlo

We implement the original, so called 'straight' version of the Event Chain Monte Carlo algorithm [44]. In the Event-Chain Monte Carlo algorithm, an activity label is assigned to a particle $i$ chosen uniformly at random, together with a direction of motion $\boldsymbol{v} \in \{\mathbf{e}_x, \mathbf{e}_y\}$. The active particle is displaced along the direction $\boldsymbol{v}$ until a collision with another particle $j$ occurs. After the collision, the activity label passes from particle $i$ to particle $j$. The latter starts then to move along the direction $\boldsymbol{v}$. The iteration of this procedure produces a driven, collective displacement of a chain of particles. When the sum of the displacements of all the particles involved in the chain add up to a value $\ell$, the activity label and direction of self propulsion are uniformly resampled, initiating a new chain. One time step of the algorithm corresponds to one collision among the hard spheres, or to the random resampling of the activity label and self-propulsion direction $\boldsymbol{v}$. We studied the Event Chain Monte Carlo dynamics for a system of $N = 1000$ polydisperse hard spheres at $\phi = 0.604$. Following [50], we chose $\ell = 0.2L$, with $L$ the linear size of the box. The relaxation curves and the time reparametrization invariant plots for Event Chain Monte Carlo, shown in the SI, have been obtained by averaging over 50 independent realizations of the dynamics of the system in the steady state.

### Swap Monte Carlo

In the Swap algorithm, one alternates between a set of $N$ Metropolis moves and $N$ swap moves. A set of $N$ Swap moves is performed with probability $p_{\mathrm{Swap}}$. During a swap move, a pair of particles is selected uniformly at random and an exchange of the particle diameters is proposed. If the exchange does not generate overlaps with the neighbors, the move is accepted. The relaxation curves displayed in Fig. 3 of the main text and in Fig. 4 of the SI have been obtained by averaging over 35 independent realizations of the Swap dynamics in the equilibrium state.

### Collective Swap

The collective Swap algorithm implementation for a three-dimensional system of polydisperse hard spheres is described in [50]. We alternate randomly between a set of $N$ Metropolis moves and a set of $N/2$ collective Swap moves. The latter set occurs with probability $p_{\mathrm{cSwap}} = 0.2$. We studied the collective Swap dynamics for s system of $N = 1000$ polydisperse hard spheres at $\phi = 0.648$. The relaxation curves and the time reparametrization invariant plots, shown in the SI, are obtained by averaging over 25 independent realizations of the dynamics in the steady state.

### East model with soft kinetic constraints and 'swap' softness updates

We consider $N$ sites on a one dimensional, periodic lattice. Each site $i$ has a spin value $n_i \in \{0, 1\}$ and a softness value $s_i \in \{0, 1\}$. The Hamiltonian $H$ of the system is the one of $2N$ non interacting spins

$$H = J \sum_i n_i + B \sum_i s_i. \tag{6}$$

The kinetic constraints of the model are implemented by means of a constraint function $C_i$ for each site:

$$C_i = n_{i-1} + s_i, \tag{7}$$

which controls the rate at which a spin in site $i$ flips. A spin $n_i$ flips from state 1 to state 0 with rate $C_i$, and from state 0 to state 1 with rate $\mathrm{e}^{-J/T}$.

The softness parameter $s_i$ can be updated in two ways [24]: by means of spontaneous fluctuations or by means of 'swaps' (s-updates). Spontaneous softness fluctuations take place only on site with $n_i = 1$, with rate $r_x = \mathrm{e}^{-J/T}$. When a spontaneous softness fluctuation occurs at site $i$ the value of $s_i$ becomes 1 or 0 with probability $(1+\mathrm{e}^{B/T})^{-1}$ or $(1+\mathrm{e}^{-B/T})^{-1}$, respectively. s-updates, on the other hand can occur on any site, independently from the value of $n_i$, with rate $r_s$, which is a parameter of the model and, when nonzero, is proportional to $\mathrm{e}^{-J/T}$. This ensures that the s-updates dynamics takes place on a similar timescale as for the creation of excitations. During an s-update, the value of the softness is updated using the same probabilities as for the spontaneous softness fluctuations.

Following [24], we fix the energy scale of the softness to $B/T = 2$. This ensures that the introduction of s-updates yields substantial speedup to the dynamics. In fact, for values of $B/T$ too low or too big the softness and the excitation dynamics decouple, making s-updates less effective.

Since the thermodynamics of the system is the one for a system of noninteracting spins, equilibrium initial condition at a given temperature can be directly generated. The dynamics of the system is instead simulated using the Botz-Kalos-Lebowitz algortihm [66], or continuous time Monte-Carlo. In a nutshell, this is a rejection-free method that relies on computing the time that the system spends in a given configuration before transitioning to a new one, instead of proposing moves toward new configurations that are prone to rejection. At each step of the algorithm, we

1. We start from the current configuration $\mathcal{C}$ of the system at time $t$, given by an assignment of the spins $n_i$ and softness $s_i$ to the $N$ sites.

2. Enumerate the $M$ configurations $\mathcal{C}_1, \ldots, \mathcal{C}_M \neq \mathcal{C}$ that the system can evolve into, starting from $\mathcal{C}$, and compute the rate $\omega_k$ at which the transition $\mathcal{C} \to \mathcal{C}_k$ can happen.

3. We compute the cumulative sum $S \equiv \sum_{k=1}^{M} \omega_k$.

4. We draw a configuration $\mathcal{C}^*$ from the set of $M$ possible configurations. the probability weight of a configuration $k$ is $\omega_k/S$.

5. We update the configuration $\mathcal{C}$ to the new configuration $\mathcal{C}^*$, and we increment the time by an amount $\Delta t = \frac{\log 1/r}{S}$, with $r$ a random number uniformly distributed in the interval $(0, 1]$.

6. We update $\mathcal{C}^* \to \mathcal{C}$, $t + \Delta t \to t$ and we start back from step 1.

The data presented in the main text in Fig. 4 are the results of an average over 50 independent runs for a system of $N = 512$ sites.

### $p$-spin with Ichiki-Ohzeki dynamics

A $p$-spin spherical glass consists of $N$ continuous spins $\sigma_i$ on a fully connected lattice, interacting through a $p$-body Hamiltonian $H(\sigma)$. A quench disorder is introduced by means of random, coupling constants among the spins.

$$H(\sigma) \equiv \sum_{i_1 < \ldots < i_p} J_{i_1 \ldots i_p} \sigma_{i_1} \ldots \sigma_{i_p}. \tag{8}$$

The coupling constants $J_{i_1 \ldots i_p}$ are independent Gaussian random variable encoding the quenched disorder of the system, with variance $\overline{\left(J_{i_1 \ldots i_p}\right)^2} = \frac{p!}{2N^{p-1}}$. The equilibrium overdamped Langevin dynamics for this model reads

$$\dot{\sigma}_i = F(\sigma_i, t) + \sqrt{2T}\xi_i(t), \tag{9}$$

where $\xi_i(t)$ is a Gaussian white noise with correlations $\langle \xi_i(t)\xi_j(t') \rangle = 2T\delta(t - t')$. The force $F_i(\sigma_i, t) = -\frac{\partial H}{\partial \sigma_i}$ contains a contribution coming from the gradient of the Hamiltonian and a harmonic restoring force, which ensures that the spherical constraint $\sum_i \langle \sigma_i(t)^2 \rangle = N$ is satisfied at all times.

The fully connected nature of the model allows to study its dynamics by means of correlations and response functions $C(t, t')$, $R(t, t')$, defined respectively as

$$\begin{aligned} R(t, t') &= \frac{1}{N} \sum_i \overline{\left\langle \left. \frac{\partial \sigma_i(t)}{\partial h_i(t')} \right|_{h_i = 0} \right\rangle} \\ C(t, t') &= \frac{1}{N} \sum_i \overline{\langle \sigma_i(t)\sigma_i(t') \rangle}. \end{aligned} \tag{10}$$

To illustrate the concept of time reparametrization invariance, we consider an alternative dynamics for the $p$-spin. It exploits the possibility of injecting a nonequilibrium drive in the system which is specifically tailored to ensure that the steady state of the system follows the Boltzmann distribution. The use of these kind of dynamics can be rewarded by faster convergence [67]. In practice, the irreversible drift is implemented by means of the so-called Ichiki-Ohzeki dynamics [68].

The nonreciprocal coupling is obtained considering two $p$-spin models, each having $N$ spins $\sigma_i^{(\alpha)}$ with independent quenched disorders and a total Hamiltonian given by the sum of the Hamiltonians of the two systems, $H_{\text{tot}}(\sigma^{(1)}, \sigma^{(2)}) = H(\sigma^{(1)}) + H(\sigma^{(2)})$. The two system evolve according to an overdamped Langevin dynamics, which contains an anti-symmetric coupling between the two copies:

$$\begin{bmatrix} \dot{\sigma}_i^{(1)}(t) \\ \dot{\sigma}_i^{(2)}(t) \end{bmatrix} = \begin{bmatrix} 1 & -\gamma \\ \gamma & 1 \end{bmatrix} \begin{bmatrix} F^{(1)}(\sigma_i^{(1)}, t) \\ F^{(2)}(\sigma_i^{(2)}, t) \end{bmatrix} + \begin{bmatrix} \sqrt{2T}\xi_i^{(1)}(t) \\ \sqrt{2T}\xi_i^{(2)}(t) \end{bmatrix}. \tag{11}$$

The realization of the Gaussian noises $\xi_i^{(\alpha)}(t)$ are independent from one system to the other. The parameter $\gamma$ encodes the strength of the nonreciprocal forces exerted between the two copies. For $\gamma = 0$, we fall back to the case of two independent $p$-spin models evolving through an equilibrium dynamics. When $\gamma \neq 0$, the dynamics becomes out of equilibrium, but it admits the Boltzmann distribution $\rho_{\text{B}} \propto \mathrm{e}^{-\beta H_{\text{tot}}}$ in its steady state, which is reached with a shorter relaxation time compared to the equilibrium case [57].

The steady state dynamics of (11) has been studied previously [57], quantifying the acceleration of the system in an ergodic region above $T_d$, the dynamical transition temperature below ergodicity is broken. $T_d$ is the same as in equilibrium. Below a critical temperature $T_d$, the ergodicity of the system is broken. Here, we investigate the role of time reparametrization invariance by considering the dynamics of the coupled systems at temperature slightly above or slightly below $T_d$, starting from an initial configuration at infinite temperature. We consider a generalization of (10) to encode responses and correlations internal to each copy and among each copy of the system in two $2 \times 2$ matrices $R_{\alpha\beta}(t, t')$ and $C_{\alpha\beta}(t, t')$. From this quantities, which encode the full time evolution of the system, we isolate a slow contribution $\widetilde{C}_{\alpha\beta}(t, t')$, $\widetilde{R}_{\alpha\beta}(t, t')$, for which

$\partial_t \widetilde{C}_{\alpha\beta}(t,t') \approx \partial_t \widetilde{R}_{\alpha\beta}(t,t') \approx 0$. Imposing the *ansatz* $C_{\alpha\beta} = \delta_{\alpha\beta}\widetilde{C}(t,t')$, $\widetilde{R}_{\alpha\beta} = \frac{1}{1+\gamma^2}\widetilde{R}(t,t')(\delta_{\alpha\beta}+\gamma\epsilon_{\alpha\beta})$, with $\epsilon_{\alpha\beta}$ the Levi-Civita tensor, we obtain the pair of integral equations

$$
\begin{aligned}
0 &= \left[-\frac{T}{1-q} + \frac{p(p-1)}{2T}(1-q)\widetilde{C}(t,t')^{p-2}\right]\widetilde{R}(t,t') + \frac{p(p-1)}{2}\int_{t'}^{t} \mathrm{d}\tau \widetilde{C}(t,\tau)^{p-2}\widetilde{R}(t,\tau)\widetilde{R}(\tau,t') \\
0 &= \left[-\frac{T}{1-q} + \frac{p}{2T}(1-q)\widetilde{C}(t,t')^{p-2}\right]\widetilde{C}(t,t') + \frac{p}{2}\int_{t'}^{t} \mathrm{d}\tau \widetilde{C}(t,t')^{p-1}\widetilde{R}(\tau,t') \\
&+ \frac{p(p-1)}{2}\int_{0}^{t} \mathrm{d}\tau \widetilde{C}(t,\tau)^{p-2}\widetilde{R}(t,\tau)\widetilde{C}(\tau,t').
\end{aligned}
\tag{12}
$$

If $T > T_d$, the quantity $q$ is the height of the transient, long-lived plateau crossed by the correlation function during the relaxation process. If $T \leq T_d$, $q$ is the Edwards-Anderson order parameter [69], which can be determined from the thermodynamics of the system. More details about the derivation can be found in [39]. These Equations define explicitly the functionals $\mathcal{F}$ and $\mathcal{G}$ introduced in (2).

# Supplementary Material for
# "Clever algorithms for glasses work by time reparametrization"

Federico Ghimenti,[1,2] Ludovic Berthier,[3,4] Jorge Kurchan,[5] and Frédéric van Wijland[1]

[1] *Laboratoire Matière et Systèmes Complexes (MSC), Université Paris Cité & CNRS (UMR 7057), 75013 Paris, France*
[2] *Department of Applied Physics, Stanford University, Stanford, CA 94305, USA*
[3] *Laboratoire Charles Coulomb (L2C), Université de Montpellier & CNRS (UMR 5221), 34095 Montpellier, France*
[4] *Gulliver, UMR CNRS 7083, ESPCI Paris, PSL Research University, 75005 Paris, France*
[5] *Laboratoire de physique LPENS, 24 rue Lhomond, 75005 Paris, France*

## I. TRANSVERSE FORCES: SELF-INTERMEDIATE SCATTERING FUNCTION

In the main text, the relaxation of the Kob-Andersen mixture has been tracked by means of the overlap function $F_o(a,t)$ for the large-diameter particles of type $A$, namely

$$F_o(a,t) \equiv \frac{1}{N_A}\sum_{i,j}\langle\theta(a - |\mathbf{r}_i^{(A)}(t) - \mathbf{r}_j^{(A)}(0) - \Delta\mathbf{r}_{com}|)\rangle, \tag{1}$$

where $\Delta\mathbf{r}_{\rm com}$ is the displacement of the center of mass of the system from time 0 to time $t$. This function was previously introduced [1] to track the collective relaxation of the system, while being less prone to noise compared to the Fourier transform of the density-density autocorrelation function. A related –and certainly more common– quantity extensively used to track the relaxation of supercooled liquids is the self part of the intermediate scattering function, $F_s$. For the binary mixture studied here $F_s$ is defined as

$$F_s(\mathbf{k},t) \equiv \frac{1}{N_A}\sum_{i}\langle \mathrm{e}^{-i\mathbf{k}\cdot\left[\mathbf{r}_i^{(A)}(t) - \mathbf{r}_i^{(A)}(0) - \Delta\mathbf{r}_{com}\right]}\rangle. \tag{2}$$

The modulus of the wave vector $\mathbf{k}$ sets the inverse length scale over which the relaxation of the system is tracked.

We demonstrate numerically that, as far as the speedup is due to a change in the translational dynamics, time reparametrization is robust against a change of observable from $F_o$ to $F_s$ (we postpone to Section III an explanation of why this is not adapted to deal with Swap dynamics and thus not convenient for comparing various dynamics). Our results are shown in Fig. 1(a) and (b), where we plot the time evolution of $F_s$, evaluated at two distinct wave vectors $\mathbf{k}_1 = \frac{2\pi}{L}(1,4,0)$ and $\mathbf{k}_2 = \frac{2\pi}{L}(0,4,1)$, for different values of the strength of transverse forces $\gamma$. For both wavevectors we have $|\mathbf{k}_1| = |\mathbf{k}_2| \approx 7.21$, corresponding to the location of the first peak of the structure factor for particles of species $A$. Since the transverse force dynamics is anisotropic, the curves for $F_s$ have different behaviors depending on the direction where the wave vector is pointing. This difference can be appreciated in the short time behavior of $F_s$. In the presence of transverse forces, the relaxation of the system is faster compared with the one observed for equilibrium dynamics by a factor $\approx 3$. Both curves exhibit signs of a two-step relaxation, and in the late stage of the relaxation the shapes of the curves are very similar, even when passing from $\gamma = 0$ to $\gamma = 4$. In Fig. 1(c), we represent $F_s(\mathbf{k}_2, t)$ as a function of $F_s(\mathbf{k}_1, t)$ in a parametric plot, for $\gamma = 0$ and $\gamma = 4$. The curves obtained in this way are invariant under a reparametrization of time. Interestingly the curve for equilibrium dynamics and for transverse force dynamics are different from each other at the beginning of the relaxation, when $F_s$ lie approximately between 1 and 0.8. As the relaxation of the system proceeds, at later times, the two curves remarkably collapse on top of each other. This result demonstrates that the time reparametrization invariance evidenced by transverse forces is robust with respect to a change of the observable used to track the relaxation of the system.

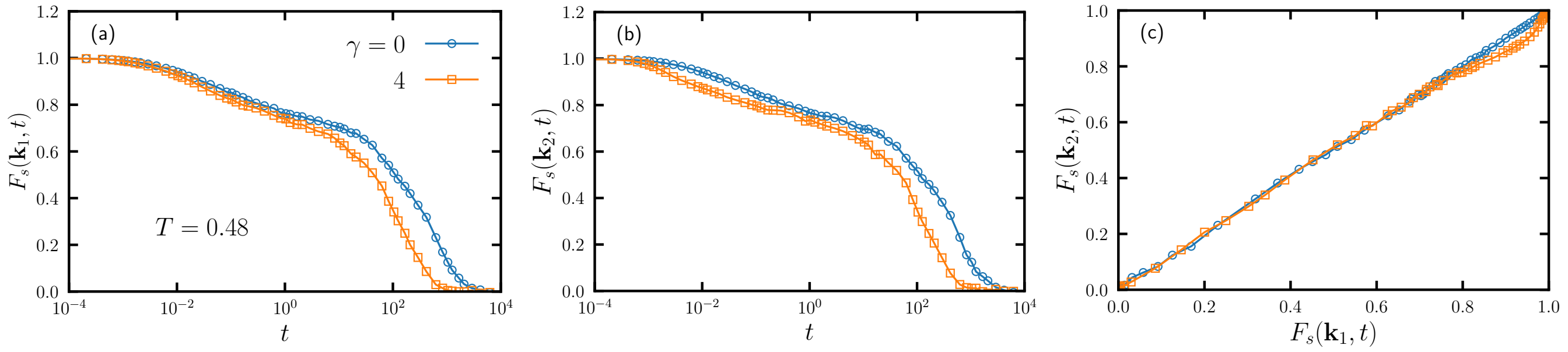

FIG. 1. (a-b) Time evolution of the self intermediate scattering function for a Kob-Andersen mixture under equilibrium overdamped Langevin dynamics ($\gamma = 0$) and with transverse forces ($\gamma = 4$), for two different wavevectors: $\mathbf{k}_1$ has a nonzero component in the direction $z$. This direction is not directly affected by transverse forces. $\mathbf{k}_2$ lies instead entirely within the $xy$ plane. (c) Time reparametrization invariant plot, obtained representing $F_s(\mathbf{k}_2,t)$ as a function of $F_s(\mathbf{k}_1,t)$. After a first fast relaxation to the plateau, the curves obtained with and without transverse forces collapse onto the same master curve.

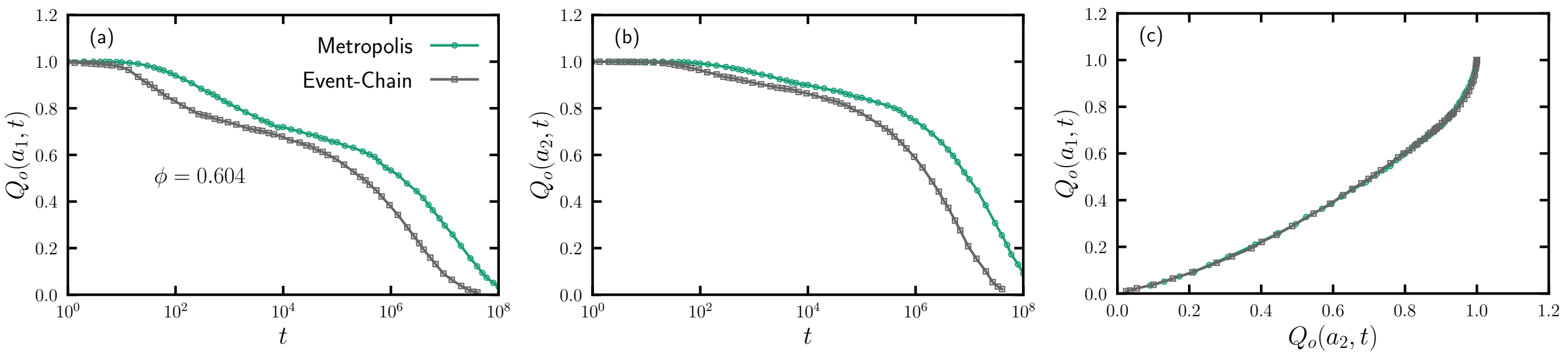

FIG. 2. (a-b) Time evolution of the collective overlap function $Q_o(a, t)$ for polydisperse hard sphere under Metropolis (green circles) and Event Chain (gray squares) dynamics. Two different cutoff lengths are used: $a_1 = 0.2\overline{d}$, $a_2 = 0.3\overline{d}$. (c) Parametric plot of $Q_o(a_1, t)$ as a function of $Q_o(a_2, t)$. The curves obtained with the Metropolis and Event Chain dynamics collapse onto the same master curve. .

## II. EVENT-CHAIN MONTE CARLO: NUMERICAL RESULTS

We implement the original, 'straight' version of the Event Chain Monte Carlo algorithm [2] for the polydisperse hard spheres described in Methods. Here we compare the dynamics of polydisperse hard spheres under Event-Chain evolution with the Metropolis one, through the lens of time reparametrization invariance ideas. We study the dynamics of $N = 1000$ polydisperse hard spheres, and we measure the time evolution of the normalized overlap function, given by (2) of the main text.

Starting from equilibrated samples at fixed density, we compare the evolution for the system under Metropolis and Even Chain Monte Carlo. We track the collective overlap function $Q_o$ using two different cutoff lengths $a_1$ and $a_2$, which are then used to construct a parametric plot that proves reparametrization invariant. Our results are shown in Fig. 2. When plotting $Q_o$ as a function of time, as done in Fig. 2(a-b), all the curves exhibit a two step relaxation. The speedup of Event Chain over Metropolis can be quantified by looking at a cutoff-dependent relaxation time $\tau_\alpha(a)$ defined by $Q_o(a, \tau_\alpha) = \mathrm{e}^{-1}$. Using this timescale, Event Chain beats the Metropolis algorithm by a factor $\approx 5$. We then show a parametric plot of one overlap function as a function of the other in Fig. 2. The collapse of the curves onto a single master curve demonstrates that Event Chain operates, in this regime, by means of a reparametrization of time.

We complement our analysis with fully consistent results for the self intermediate scattering function $F_s(\mathbf{k}, t)$, defined as

$$F_s(\mathbf{k}, t) \equiv \frac{1}{N} \sum_i \langle \mathrm{e}^{-i\mathbf{k}\cdot\Delta\mathbf{r}_i(t)} \rangle, \tag{3}$$

with $\Delta\mathbf{r}_i(t) \equiv \mathbf{r}_i(t) - \mathbf{r}_i(0) - \Delta\mathbf{r}_{\mathrm{com}}(t)$ the displacement of particle $i$ relative to the displacement of the center of mass of the system, $\Delta\mathbf{r}_{\mathrm{com}}(t)$, at time $t$. We choose two values of the modulus of $\mathbf{k}$, $k_1 = \frac{2\pi}{\overline{d}}$, $k_2 = \frac{4\pi}{\overline{d}}$. For each of these values $k_i$, we average our data for $F_s$ over all the wave vectors $\mathbf{k}$ such that $|\mathbf{k}| = k_i$, thus obtaining curves for two different correlation functions $F_s(k_1, t)$, $F_s(k_2, t)$. These two functions track single particle motion over length scales corresponding to a particle diameter and half a diameter, respectively. The behavior of these curves as a function of time for a fixed density for Event-Chain and the Metropolis algorithm is shown in Fig. 3(a-b). For a fixed value of $k$, the curves feature a two-step relaxation with both dynamics. As $k$ increases, the height of the plateau decreases. This is because by increaseing $k$ we probe the decorrelation of density fluctuations on a shorter length scale, where for a fixed time a larger number of particles have relaxed. For both values of $k$, we see that the relaxation is faster under Event-Chain dynamics by about an order of magnitude. However, the shape of the curves under Event-Chain dynamics is visually similar to the one obtained under Metropolis dynamics. With Event-Chain Monte Carlo, a plateau of a height comparable to the one observed for Metropolis dynamics is reached at shorter times. In Fig. 3(c) we show the parametric plot that can be obtained from the two intermediate scattering functions. In the parametric plots, the curves obtained with the Metropolis and Event Chain algorithm are identical. These results demonstrate that the Event Chain Monte Carlo algorithm is identical to Metropolis Monte Carlo in glassy systems up to a reparametrization of time.

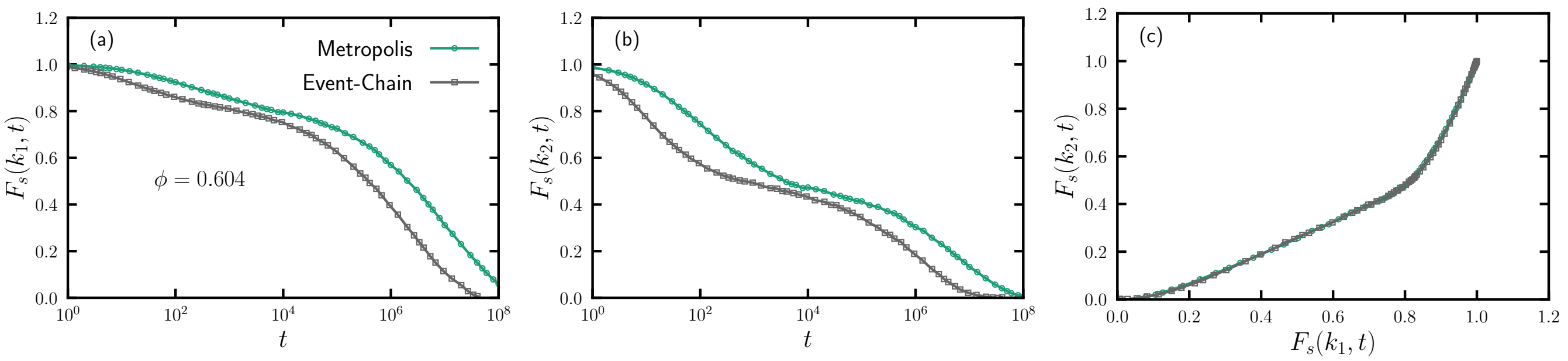

FIG. 3. (a-b): Time evolution of the self intermediate scattering function for the Metropolis and the Event Chain Monte Carlo algorithms in a three dimensional polydisperse hard sphere system of $N = 1000$ particles. The moduli of the wavevectors are $k_1 = \frac{2\pi}{d}$, $k_2 = \frac{4\pi}{d}$. (c): Parametric plots of $F_s(\mathbf{k}_2, t)$ under Metropolis and Event-Chain Monte Carlo as a function of $F_s(\mathbf{k}_1, t)$ for the three possible combinations of wavevectors. The curves collapse onto the same master curve.

## III. WHAT TO LOOK AT: SWAP ALGORITHMS AND SELF INTERMEDIATE SCATTERING FUNCTION

In Sections I and II we have shown that time reparametrization invariance is insensitive to whether the relaxation is tracked using collective quantities – such as the overlap function $F_o$ – or single-particle quantities – such as the self-intermediate scattering function $F_s$. This is true as far as only translational moves are involved. However, when comparing, in the parametric plots, swap algorithms with their swap-free counterparts, there is a caveat in resorting to observables related to single-particle quantities. This is because $F_s$ treats a *spontaneous* exchange of particle position produced by diffusive motion in position space as a form of relaxation. On the other hand swap moves, which we are implemented by exchanging the particle diameters, practically achieve the same result but do not decrease the value of $F_s$. This issue disappears from functions that track collective relaxation such as $Q_o$, hence or choice to focus on $Q_o$ for all algorithms.

This observation has a qualitative consequence on the relative shapes of the parametric plots obtained using two self intermediate scattering functions $F_s(k_1, t)$, $F_s(k_2, t)$ for swap and swap-free algorithms. Let us take $k_2 > k_1$, so that $F_s(k_2, t)$ tracks the motion of the particles over shorter length scales than $F_s(k_1, t)$. In the absence of swap moves, a spontaneous exchange of the particle positions over a length scale larger than $\sim k_1^{-1}$ necessarily leads to a decrease of both $F_s(k_1, t)$ and $F_s(k_2, t)$. On the other hand, a swap move between particles separated by a length scale $\sim k_1^{-1}$ followed by a particle displacement over the shorter scale $\sim k_2^{-1}$ produces a decrease in $F_s(k_2, t)$, but not in $F_s(k_1, t)$. This means that on the parametric plot, we expect the swap curves to bend towards the $F_s(k_1, t)$ axis. This is indeed observed numerically in Fig. 4.

## IV. COMPARISON OF SWAP DYNAMICS AT FIXED SPEEDUP AND DIFFERENT DENSITIES

As stated in the main text, we expect time reparametrization invariance to hold among two algorithms as long as they operate in a regime with slow, two step relaxation, and a well defined plateau, at the same height for both the dynamics. For swap algorithm, approaching this regime is challenging since the speedup provided at fixed swap probability $p_s$ increases with increasing density or lowering the temperature [1]. An alternative route consists in comparing parametric plots for swap and Metropolis dynamics different densities, and different values of the swap probabilities for each density. These values are to be chosen so that the speedup of Swap over Metropolis remains constant as the density increases. A comparison of this type is performed in Fig. 5, where the swap probability is adjusted as the density increases so that a constant speedup of $\approx 5$ is maintained. A slight reduction of the discrepancy among the two curves is observed.

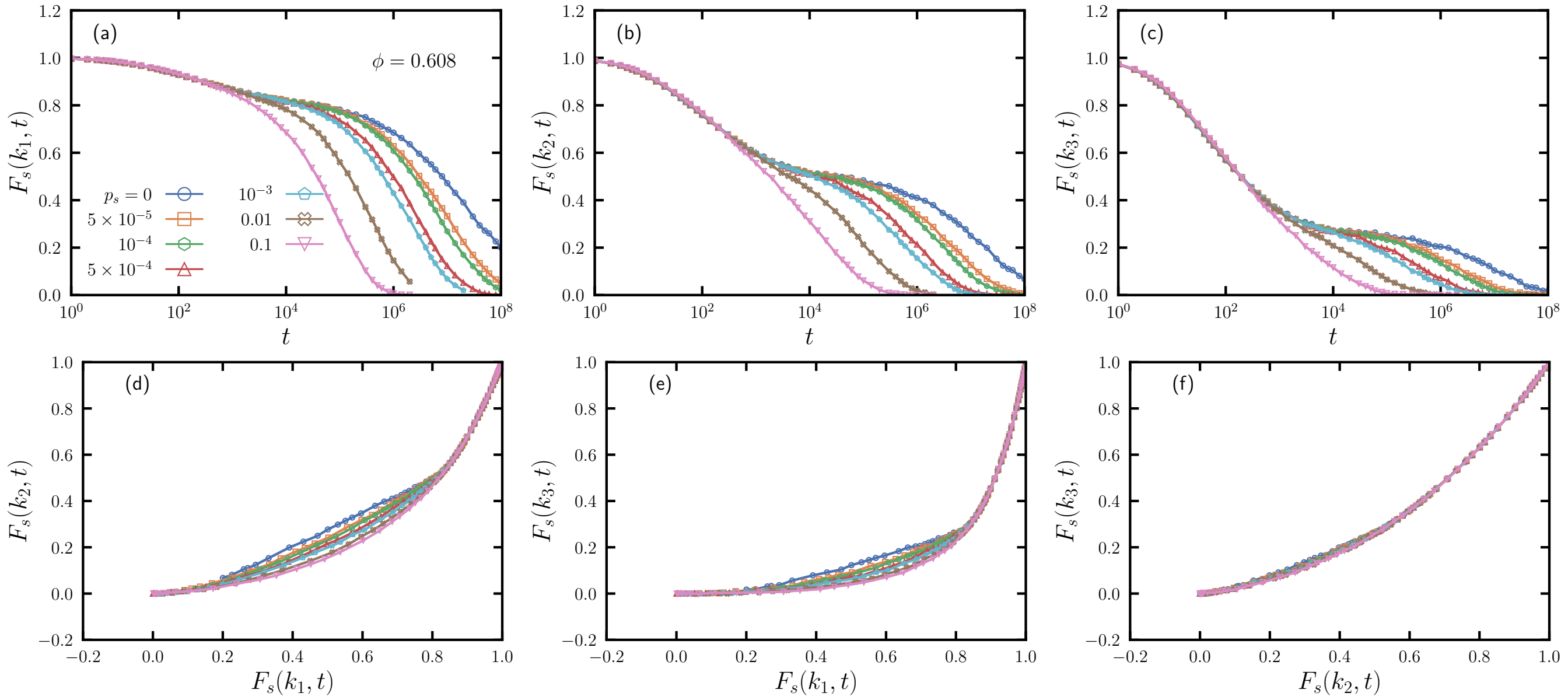

FIG. 4. (a-c): Time evolution of the self intermediate scattering function for the Metropolis (blue circles) and Swap Monte Carlo algorithms with different swap probabilities in a three dimensional polydisperse hard sphere system of $N = 1000$ particles. The three panels correspond to three different wave vector moduli, $k_1 = \frac{2\pi}{\overline{d}}$, $k_2 = \frac{4\pi}{\overline{d}}$, $k_3 = \frac{6\pi}{\overline{d}}$. (d-f): Parametric plots of $F_s(k_i, t)$ under Metropolis and Swap Monte Carlo as a function of $F_s(k_j, t)$ for the three possible combinations of wave vectors. The collapse is imperfect, which is a consequence of using single-particle functions to track the relaxation. Consistently with our discussion in Sec. III, in the parametric plots the swap curve bends towards the axis related to displacement over the longest length scales.

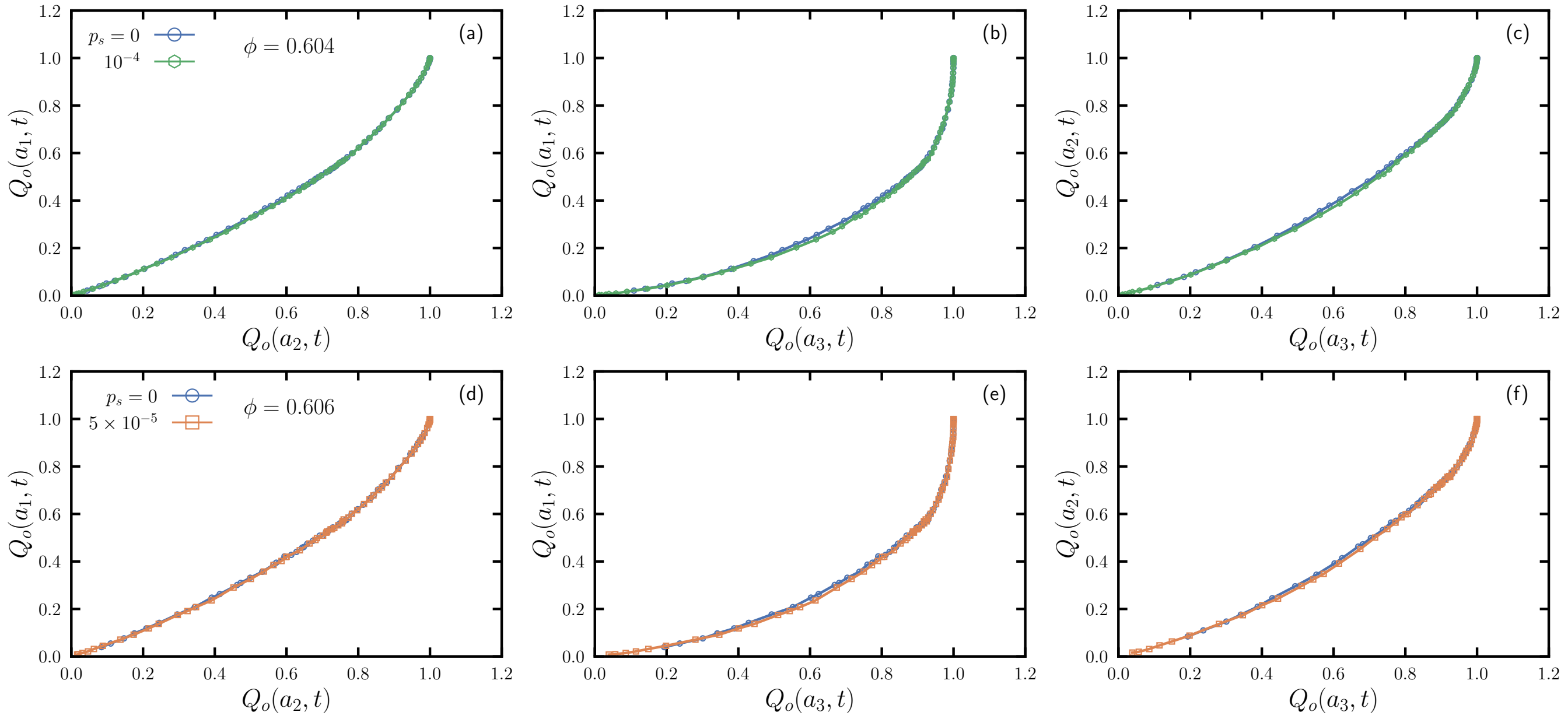

FIG. 5. Parametric plots of collective overlap functions at different values of the cutoff distance $a$, comparing Metropolis and Swap algorithm at two different densities, with the swap probabilities adjusted as the density increases to preserve a constant speedup of $\approx 5$. In each panel, a different value of $a$ is used. $a_1 = 0.15\overline{d}$, $a_2 = 0.2\overline{d}$, $a_3 = 0.3\overline{d}$. (a-c): parametric plots obtained using different combinations of $Q_o(a_i, t)$ and $Q_o(a_j, t)$ for Metropolis against Swap at $\phi = 0.604$ and $p_s = 10^{-4}$. (e-f): parametric plots obtained using different combinations of $Q_o(a_i, t)$ and $Q_o(a_j, t)$ for Metropolis against Swap at $\phi = 0.606$ and $p_s = 5 \times 10^{-5}$ at $\phi = 0.606$.

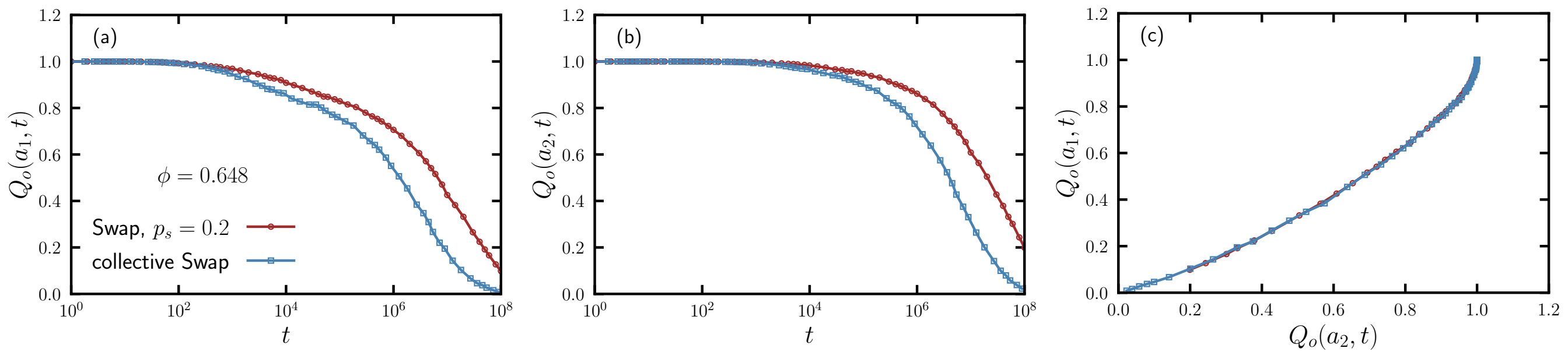

FIG. 6. Time evolution of the collective overlap function $Q_o(a, t)$ for polydisperse hard sphere under Swap (brown circles) and collective Swap (blue squares) dynamics. Two different cutoff lengths are used: $a_1 = 0.2\bar{d}$, $a_2 = 0.3\bar{d}$. (c) Parametric plot of $Q_o(a_1, t)$ as a function of $Q_o(a_2, t)$. The curves obtained with the Metropolis and collective Swap dynamics collapse onto the same master curve. .

## V. COLLECTIVE SWAPS

In this Section we compare the Swap and collective Swap dynamics. For the implementation of the collective Swap algorithm and the parameter used in this study, we refer the reader to [3]. In Fig. 6 (a-b) we show the time evolution of the collective overlap function $Q_o(a, t)$ for Swap and collective Swap dynamics. The two panels correspond respectively to two values of the cutoff length $a$, as carried out in the study of the Event Chain Monte Carlo algorithm in Sec. II. The collective Swap dynamics allows for a faster relaxation of $Q_o$, by a factor $\approx 5$. In Fig. 6 (c) we show parametric plots of $Q_o(a_1, t)$ *versus* $Q_o(a_2, t)$ . Within this representation, the curves obtained with Swap and collective Swap collapse onto of each other, demonstrating that collective swaps too accelerate the dynamics in a fashion captured by time reparametrization.

## VI. A KINETICALLY CONSTRAINED MODEL WITH COLLECTIVE RELAXATION FUNCTIONS

In the main text, we presented results for soft-East model with softness updates. The relaxation of this system can only be tracked by means of single-spin relaxation functions: detailed balance and the directed nature of the kinetic constraint make the time relaxation of collective dynamical correlation functions identical to the ones of individual spins [4], and the relaxation observed is insensitive to wavevectors or lengthscales To our knowledge, this is the only kinetically constrained model known in the literature where a dynamics mimicking Swap has been implemented.

Here, we expand the scope of our investigations by reporting results for a one-dimensional soft Fredrickson-Andersen model [5–7] with softness update. The model consists of a one-dimensional periodic chain of $N$ sites. Each site $i$ is characterized by a binary variable $n_i \in \{0, 1\}$ determining whether it hosts an excitation or not, and a softness degree of freedom $s_i \in \{0, 1\}$. The Hamiltonian of the model $H^{\rm FA}$ has the same form as the one of the soft-East model, that is,

$$H^{\rm FA} = J \sum_{i=1}^{N} n_i + B \sum_{i=1}^{N} s_i \, . \tag{4}$$

The difference with the East model lies in the definition of the rates $C_i^{\rm FA} e^{-\frac{J}{T} n_i}$ at which the variables $n_i$ are updated for a given temperature $T$. In the Fredrickson-Andersen model, $C_i^{\rm FA}$ is defined as

$$C_i^{\rm FA} = n_{i+1} + n_{i-1} + s_i \, , \tag{5}$$

which means that in the absence of the softness degrees of freedom, an excitation can be created, or destroyed, at a given site if at least one of its neighbors is hosting an excitation. The update of the softness degrees of freedom takes place with the same rules as the soft East model described in the Methods section of the main text. The nondirected nature of the kinetic constraint allows to track the collective relaxation of the system by means of a normalized and connected coherent correlation function $C(k, t)$, defined as

$$C(k, t) \equiv \frac{\langle n(k, t)^* n(k, 0) \rangle - \langle n(k, t)^* \rangle \langle n(k, 0) \rangle}{\langle |n(k, 0)|^2 \rangle - |\langle n(k, 0) \rangle|^2} \, , \tag{6}$$

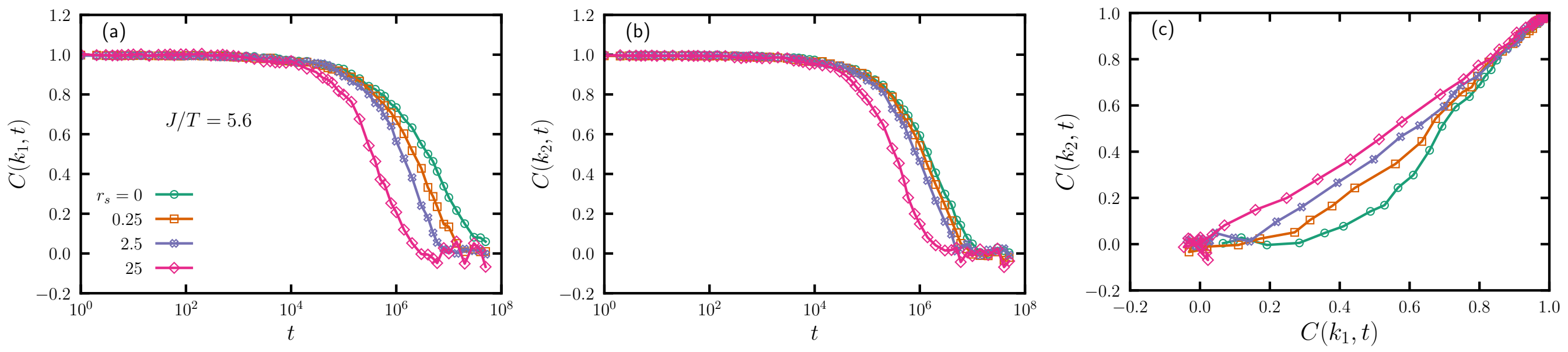

FIG. 7. (a-b) Time evolution of the collective correlation function $C(k,t)$ for a one-dimensional, soft, Fredrickson-Andersen model with softness updates at rate $r_s$. Two different wavevectors are used. In (a), we take $k_1 = 2\pi/2048$, while in (b) we take $k_2 = 2\pi/512$. (c) Parametric plot of $C(k_2,t)$ as a function of $C(k_1,t)$. An evident discrepancy between the different curves can be observed in the slow relaxation regime. .

where $n(k,t) \equiv \frac{1}{\sqrt{N}} \sum_{j=1}^{N} \mathrm{e}^{ikj} n_j$ is the Fourier transform of the field of excitations. The brackets $\langle \ldots \rangle$ denote an average over both the Boltzmann distribution for the initial conditions and realizations of the dynamics. For a lattice of size $N$, the wavevector $k$ can take the values $\frac{2\pi}{N} j$, with $j$ an integer.

We simulate the dynamics of a system composed of $N = 4096$ spins, with $J/T = 5.6$ and $B/T = 2J/T$, using the continuous time Monte Carlo algorithm described in Methods for different values of the softness update rate $r_s$. For each value of $r_s$ we perform up to 2400 independent runs starting from independent equilibrium initial conditions. From these runs, the time evolution of the collective correlation function $C(k,t)$, defined in (6), is measured for different values of the wavevector $k$. Our results are shown in Fig. 7. In absence of softness updates, the relaxation of the system unfolds over 7 decades in time. Enabling softness updates yields a speedup up to an order of magnitude, but we also observe that the speedup changes depending on the lengthscale over which the relaxation is tracked. A qualitative explanation of this trend is given by observing that, in the stationary state, the probability to find a soft site in a portion of the system of size $k^{-1}$ decreases as $k$ increases. The parametric plot of one correlation function for a given wavelength against the other shows absence of collapse, similarly to the result presented in the main text for a different kinetically constrained model and for different dynamical observables. These results suggest that even when collective variables are studied, kinetically constrained models lack time reparametrisation invariance.

The space of possible kinetic constraints and lattice geometries that can be designed is, of course, quite large. It would be certainly interesting to extend the dynamics with spontaneous softness updates to other kinetically constrained models in two and three dimensions [8] or on the Bethe lattice [9], and to probe time reparametrization in these contexts too.

## VII. TIME REPARAMETRIZATION INVARIANCE FOR THE $p$-SPIN UNDER ICHIKI-OHZEKI DYNAMICS

A $p$-spin spherical glass consists of $N$ spins $\sigma_i$ on a fully connected lattice, interacting through a $p$-body Hamiltonian $H(\sigma)$. A quench disorder is introduced by means of random, coupling constants among the spins.

$$H(\sigma) \equiv \sum_{i_1 < \ldots < i_p} J_{i_1 \ldots i_p} \sigma_{i_1} \ldots \sigma_{i_p}. \tag{7}$$

The coupling constants $J_{i_1 \ldots i_p}$ are independent Gaussian random variable encoding the quenched disorder of the system, with variance $\overline{(J_{i_1 \ldots i_p})^2} = \frac{p!}{2N^{p-1}}$. The equilibrium overdamped Langevin dynamics for this model reads

$$\dot{\sigma}_i = F(\sigma_i, t) + \sqrt{2T} \xi_i(t), \tag{8}$$

where $\xi_i(t)$ is a Gaussian white noise with correlations $\langle \xi_i(t) \xi_j(t') \rangle = 2T\delta(t-t')$. The force $F_i(\sigma_i, t) = -\frac{\partial H}{\partial \sigma_i} + \mu(t)\sigma_i$ contains a contribution coming from the gradient of the Hamiltonian and a harmonic restoring force $\mu(t)\sigma_i$. The Lagrange multiplier $\mu(t)$ ensures that the spherical constraint $\sum_i \langle \sigma_i(t)^2 \rangle = N$ is satisfied at all times.

The fully connected nature of the model allows to study its dynamics by means of correlations and response functions

$C(t,t')$, $R(t,t')$, defined respectively as

$$R(t,t') = \frac{1}{N}\sum_i \overline{\left\langle \frac{\partial \sigma_i(t)}{\partial h_i(t')}\Big|_{h_i=0} \right\rangle}$$
$$C(t,t') = \frac{1}{N}\sum_i \overline{\langle \sigma_i(t)\sigma_i(t')\rangle}. \tag{9}$$

We now consider the Ichiki-Ohzeki irreversible dynamics of two coupled copies of the $p$-spin model, as described by (12) in Methods, which we repeat here for convenience:

$$\begin{bmatrix} \dot{\sigma}_i^{(1)}(t) \\ \dot{\sigma}_i^{(2)}(t) \end{bmatrix} = \begin{bmatrix} 1 & -\gamma \\ \gamma & 1 \end{bmatrix} \begin{bmatrix} F^{(1)}(\sigma_i^{(1)},t) \\ F^{(2)}(\sigma_i^{(2)},t) \end{bmatrix} + \begin{bmatrix} \sqrt{2T}\xi_i^{(1)}(t) \\ \sqrt{2T}\xi_i^{(2)}(t) \end{bmatrix}. \tag{10}$$

Our focus is on the transient dynamics at temperature $T$, starting from an initial configuration at infinite temperature. This temperature is taken to be slightly above or slightly below $T_d$. Our analysis holds in both cases, although the interpretation of the transient behhavior changes depending wether $T > T_d$ or $T < T_d$. The dynamics of the system is entirely characterized by its $2 \times 2$ response and correlation matrices, defined respectively as

$$R_{ab}(t,t') = \frac{1}{N}\sum_i \overline{\langle \frac{\partial \sigma^{(a)}(t)}{\partial h^{(b)}(t')}\Big|_{h^{(b)}=0} \rangle}$$
$$C_{ab}(t,t') = \frac{1}{N}\sum_i \overline{\langle \sigma_i^{(a)}(t)\sigma_i^{(b)}(t')\rangle}. \tag{11}$$

Where $h_i^{(a)}(t')$ is a small force field that perturbs the dynamics of spin $i$ in system $a$. Using path integrals techniques as in [10], and exploiting the mean field nature of the model, equations of motion for the response and correlation matrices read

$$\partial_t \mathbf{C}(t,t') = -(\mathbf{1}+\mathbf{\Gamma})\boldsymbol{\mu}(t)\mathbf{C}(t,t') + \int_0^t \mathrm{d}s \mathbf{\Sigma}_R(t,s)\mathbf{C}(s,t') + 2T\mathbf{R}^T(t',t)$$
$$+ \int_0^{t'} \mathrm{d}s \mathbf{\Sigma}_C(t,s)\mathbf{R}^T(t',s) \tag{12}$$
$$\partial_t \mathbf{R}(t,t') = -(\mathbf{1}+\mathbf{\Gamma})\boldsymbol{\mu}(t)\mathbf{R}(t,t') + \int_{t'}^t \mathrm{d}s \mathbf{\Sigma}_R(t,s)\mathbf{R}(s,t') + \mathbf{1}\delta(t-t').$$

These equations are supplemented with the boundary conditions $\lim_{t'\to t}\mathbf{R}(t,t') = \mathbf{1}$ and with the spherical constraint $\mathbf{C}(t,t) = \mathbf{1}$. The matrix $\mathbf{1}$ is the $2 \times 2$ identity matrix, while the matrix $\mathbf{\Gamma} \equiv \begin{bmatrix} 0 & -\gamma \\ \gamma & 0 \end{bmatrix}$ encodes the antisymmetric coupling between the two copies. The complexity of the Markovian dynamics of this system of $2N$ spins has been reduced to a smaller set of equations, at the price of introducing memory effects through the kernels $\mathbf{\Sigma}_R(t,t')$, $\mathbf{\Sigma}_C(t,t')$. As customary in mean field approaches, the expression of these kernels is self consistently determined as a function of $\mathbf{C}(t,t')$ and $\mathbf{R}(t,t')$. Their expressions are

$$\mathbf{\Sigma}_R(t,t') = \frac{p(p-1)}{2} \times$$
$$\begin{bmatrix} C_{11}^{p-2}(R_{11}+\gamma R_{12}) - \gamma C_{21}^{p-2}(R_{21}+\gamma R_{22}) & C_{12}^{p-2}(R_{12}-\gamma R_{11}) - \gamma C_{22}^{p-2}(R_{22}-\gamma R_{21}) \\ C_{21}^{p-2}(R_{21}+\gamma R_{22}) + \gamma C_{11}^{p-2}(R_{11}+\gamma R_{12}) & C_{22}^{p-2}(R_{22}-\gamma R_{21}) + \gamma C_{12}^{p-2}(R_{12}-\gamma R_{11}) \end{bmatrix} \tag{13}$$
$$\mathbf{\Sigma}_C(t,t') \equiv (\mathbf{1}+\mathbf{\Gamma})\,\mathbf{C}_p(t,t')\left(\mathbf{1}+\mathbf{\Gamma}^T\right)$$
$$C_p(t,t')_{ij} = \frac{p}{2}C_{ij}^{p-1}(t,t').$$

Note that we omitted the $(t,t')$ dependence of the terms on the right hand side of the first equation. The matrix $\boldsymbol{\mu}(t)$ enforces the spherical constraint on the system. Its expression is

$$\boldsymbol{\mu}(t) = \mathrm{diag}\left[T\mathbf{1}_2 + \int_0^t \mathrm{d}\tau \mathbf{\Sigma}_R(t,\tau)\mathbf{C}^T(t,\tau) + \int_0^t d\tau \mathbf{\Sigma}_C(t,\tau)\mathbf{R}^T(t,\tau)\right], \tag{14}$$

where for any matrix $\mathbf{X}$ we define its diagonal part as $\text{diag}[\mathbf{X}]_{ij} \equiv \delta_{ij}X_{ij}$. Above $T_d$ in the steady state, these response and correlation matrices are time translational invariant, *i. e.* $\mathbf{C}(t,t') = \mathbf{C}(t-t')$, and they obey an *accidental fluctuation-dissipation theorem* (AFDT) [10]:

$$-\frac{1}{T}\partial_t \mathbf{C}(t) = \mathbf{R}(t)(\mathbf{1}+\mathbf{\Gamma}). \tag{15}$$

Through the AFDT, a relation between the correlation and response memory kernels holds,

$$-\frac{1}{T}\partial_t \mathbf{\Sigma}_C(t) = \mathbf{\Sigma}_R(t)(\mathbf{1}+\mathbf{\Gamma}^T), \tag{16}$$

which will soon prove useful. In this work we focus on long waiting times, *i. e.* $t,t' \gg 1$. In such a regime, we assume that the *weak ergodicity breaking hypothesis* [11, 12] holds. This amounts to considering two widely separated sectors of the $t-t'$ plane. A first sector contains the domain where $\tau \equiv t-t' \ll t$. It comprises the fast decay of the correlations and the response functions, for which we assume that time translational invariance holds: $\mathbf{C}(t,t') = \mathbf{C}_f(\tau)$, $\mathbf{R}(t,t') = \mathbf{R}_f(\tau)$. Moreover, we also assume the AFDT to be valid in this regime,

$$\begin{aligned} -\frac{1}{T}\partial_\tau \mathbf{C}_f(\tau) &= \mathbf{R}_f(\tau)(\mathbf{1}+\mathbf{\Gamma}), \\ -\frac{1}{T}\partial_\tau \mathbf{\Sigma}_{C,f}(\tau) &= \mathbf{\Sigma}_R(\tau)(\mathbf{1}+\mathbf{\Gamma}^T). \end{aligned} \tag{17}$$

Using the results obtained for the steady state dynamics in [10], we assume that the correlation decays quickly to a plateau value, $\lim_{\tau\to+\infty} \mathbf{C}_f(\tau) = q\mathbf{1}$, $\lim_{\tau\to+\infty} \mathbf{R}_f(\tau) = \mathbf{0}$. If $T > T_d$, the dynamics of the system is ergodic, and $q$ is the height of the transient, long-lived plateau that the system passes through when relaxing in its stationary state. We assume it to be independent from $\gamma$. When $T \le T_d$, $q$ is instead the Edwards-Anderson order parameter [13], which is fully determined by the landscape of the model. This defines the first time sector for the relaxation of the system.

The second time sector of interest is obtained when $\frac{t-t'}{t} \sim O(1)$. In this regime, time translational invariance is broken, and the correlation and response matrices contain an aging term $\mathbf{C}(t,t') = \widetilde{\mathbf{C}}(t,t')$, $\mathbf{R}(t,t') = \widetilde{\mathbf{R}}(t,t')$. We assume these terms to vary slowly over times much larger than the times over which the dynamics is observed, $\partial_t \widetilde{\mathbf{C}}(t,t') \approx \partial_t \widetilde{\mathbf{R}}(t,t') \approx \mathbf{0}$. We further impose the following structure

$$\begin{aligned} \widetilde{\mathbf{C}}(t,t') &= \widetilde{C}(t,t')\mathbf{1} \\ \widetilde{\mathbf{R}}(t,t') &= \frac{1}{1+\gamma^2}\widetilde{R}(t,t')(\mathbf{1}+\mathbf{\Gamma}^T). \end{aligned} \tag{18}$$

We thus suppose that in the aging regime cross correlations among the two copies have vanished, but not their cross responses. Furthermore, the assumption of (18) means that the correlation and response matrices are determined by only two scalar quantities , $\widetilde{C}(t,t')$ and $\widetilde{R}(t,t')$. As a consequence, the memory kernels in the aging regime, $\widetilde{\mathbf{\Sigma}}_R(t,t')$ and $\widetilde{\mathbf{\Sigma}}_C(t,t')$, become

$$\begin{aligned} \widetilde{\mathbf{\Sigma}}_R(t,t') &= \frac{p(p-1)}{2}(\mathbf{1}+\mathbf{\Gamma})\widetilde{C}(t,t')^{p-2}\widetilde{R}(t,t') \\ \widetilde{\mathbf{\Sigma}}_C(t,t') &= \mathbf{1}(1+\gamma^2)\frac{p}{2}\widetilde{C}(t,t')^{p-1}. \end{aligned} \tag{19}$$

We will now show that the equations that determines $\widetilde{C}(t,t')$ and $\widetilde{R}(t,t')$ do not depend on $\gamma$. In particular they are identical to the ones obtained for $\gamma = 0$. We start by considering the equation for $\mathbf{C}$ in the first sector, where the evolution is fast. We have

$$\partial_\tau \mathbf{C}_f(\tau) = -(\mathbf{1}+\mathbf{\Gamma})\boldsymbol{\mu}_\infty \mathbf{C}_f(\tau) + \int_0^{t'+\tau} \mathrm{d}s \mathbf{\Sigma}_R(t,s)\mathbf{C}(s,t') + \int_0^{t'} \mathrm{d}s \mathbf{\Sigma}_C(t,s)\mathbf{R}^T(t',s). \tag{20}$$

We will now decompose the time integrals into their fast and slow regime contributions. First we introduce the quantity $t'_-$, which is used to discriminate between the two regimes. Time integrals of the form $\int_s^{t'_-} \mathrm{d}s'$, with $s < t'_-$ concern the slow regime, while time integrals of the form $\int_{t'_-}^{s} \mathrm{d}s'$ concern the fast regime. Such a separation is justified by the assumption that the two regimes are widely separated in time by the long-lived plateau of the correlation function. The separation reads:

$$\int_0^{t'+\tau} \mathrm{d}s \boldsymbol{\Sigma}_R(t,s)\mathbf{C}(s,t') \approx \int_0^{t'_-} \mathrm{d}s \widetilde{\boldsymbol{\Sigma}}_R(t,s)\widetilde{\mathbf{C}}(s,t') + \int_{t'_-}^{t'+\tau} \mathrm{d}s \boldsymbol{\Sigma}_{R,f}(t,s)\mathbf{C}_f(s,t'). \tag{21}$$

Due to the slow evolution of the aging variable, we can replace $t'_-$ with $t'$ in the upper limit of the integral $\int_0^{t'_-} \mathrm{d}s \widetilde{\boldsymbol{\Sigma}}_R(t,s)\widetilde{\mathbf{C}}(s,t')$. The second integral on the right hand side becomes, using the AFDT of (16), the definition of $\boldsymbol{\Sigma}_C$ from (13), and integration by parts:

$$\begin{aligned}
\int_{t'_-}^{t'+\tau} \mathrm{d}s \boldsymbol{\Sigma}_{R,f}(t,s)\mathbf{C}_f(s,t') &= \frac{1}{T}(\mathbf{1}+\boldsymbol{\Gamma}) \int_{t'_-}^{t'+\tau} \mathrm{d}s \partial_s \mathbf{C}_{p,f}(t-s)\mathbf{C}_f(s-t') \\
&= \frac{1}{T}(\mathbf{1}+\boldsymbol{\Gamma}) \int_{t'_-}^{t'} \mathrm{d}s \partial_s \mathbf{C}_{p,f}(t-s)\mathbf{C}_f(s-t') + \frac{1}{T}(\mathbf{1}+\boldsymbol{\Gamma}) \int_{t'}^{\tau} \mathrm{d}s \partial_s \mathbf{C}_{p,f}(t-s)\mathbf{C}_f(s-t') \\
&= \frac{p}{2T}(\mathbf{1}+\boldsymbol{\Gamma}) \left[\mathbf{C}_f(\tau) - q^p\right] - \frac{1}{T}(\mathbf{1}+\boldsymbol{\Gamma}) \int_{t'_-}^{t'} \mathrm{d}s \mathbf{C}_{p,f}(t+\tau-s)\partial_s \mathbf{C}_f(s-t) \\
&\quad - \frac{1}{T}(\mathbf{1}+\boldsymbol{\Gamma}) \int_0^{\tau} \mathrm{d}s \mathbf{C}_{p,f}(\tau-s)\partial_s \mathbf{C}_f(s).
\end{aligned} \tag{22}$$

Similarly, for the second integral we obtain

$$\begin{aligned}
\int_0^{t'} \mathrm{d}s \boldsymbol{\Sigma}_C(t,s)\mathbf{R}^T(t',s) &= \int_0^{t'_-} \mathrm{d}s \widetilde{\boldsymbol{\Sigma}}_C(t,s)\widetilde{\mathbf{R}}^T(t',s) + \int_{t'_-}^{t'} \mathrm{d}s \boldsymbol{\Sigma}_{C,f}(t-s)\mathbf{R}_f^T(t'-s) \\
&= \int_0^{t'} \mathrm{d}s \widetilde{\boldsymbol{\Sigma}}_C(t,s)\widetilde{\mathbf{R}}^T(t',s) \\
&\quad + \frac{1}{T}(\mathbf{1}+\boldsymbol{\Gamma}) \int_{t'_-}^{t'} \mathrm{d}s \mathbf{C}_{p,f}(t-s)\partial_s \mathbf{C}_f(t'-s),
\end{aligned} \tag{23}$$

while the Lagrange multiplier at long times becomes

$$\boldsymbol{\mu}_\infty \approx \mathrm{diag}\left[T\mathbf{1} + \int_0^t \mathrm{d}s \widetilde{\boldsymbol{\Sigma}}_C(t,s)\widetilde{\mathbf{R}}^T(t,s) + \int_0^t \mathrm{d}s \widetilde{\boldsymbol{\Sigma}}_R(t,s)\widetilde{\mathbf{C}}(s,t) + \frac{p}{2T}(\mathbf{1}+\boldsymbol{\Gamma})(1-q^p)\right]. \tag{24}$$

The equation of motion for the fast evolution of the correlation matrix is therefore

$$\begin{aligned}
\partial_\tau \mathbf{C}_f(\tau) &= -(\mathbf{1}+\boldsymbol{\Gamma})\boldsymbol{\mu}_\infty \mathbf{C}_f(\tau) + \frac{p}{2T}(\mathbf{1}+\boldsymbol{\Gamma}) \left[\mathbf{C}_f(\tau) - \mathbf{1}\right] \\
&\quad - \frac{1}{T}(\mathbf{1}+\boldsymbol{\Gamma}) \int_0^{\tau} \mathrm{d}s \mathbf{C}_{p,f}(\tau-s)\partial_\tau \mathbf{C}_f(s) \\
&\quad + \lim_{t\to+\infty} \left[\int_0^t \mathrm{d}s \widetilde{\boldsymbol{\Sigma}}_R(t,s)\widetilde{\mathbf{C}}(s,t) + \int_0^t \mathrm{d}s \widetilde{\boldsymbol{\Sigma}}_C(t,s)\widetilde{\mathbf{R}}^T(t,s)\right] + \frac{p}{2T}(\mathbf{1}+\boldsymbol{\Gamma})[1-q^p].
\end{aligned} \tag{25}$$

(24) and (25) describe the decay of the correlation matrix to its plateau value $q\mathbf{1}$. Note that the dynamics depends also on the slow contribution to correlation and responses.

We now write the equations that determine the slow functions $\widetilde{\mathbf{C}}(t,t')$ and $\widetilde{\mathbf{R}}(t,t')$:

$$\begin{aligned}
0 &= (\mathbf{1}+\boldsymbol{\Gamma}) \left[-\boldsymbol{\mu}_\infty + \frac{p}{2T}(1-q^{p-1})\right] \widetilde{\mathbf{R}}(t,t') + \frac{1}{T(1+\gamma^2)} \widetilde{\boldsymbol{\Sigma}}_R(t,t')(1-q)(\mathbf{1}+\boldsymbol{\Gamma}^T) \\
&\quad + \int_{t'}^t \mathrm{d}s \widetilde{\boldsymbol{\Sigma}}_R(t,s)\widetilde{\mathbf{R}}(s,t') \\
0 &= (\mathbf{1}+\boldsymbol{\Gamma}) \left[-\boldsymbol{\mu}_\infty + \frac{p}{2T}(1-q^{p-1})\right] \widetilde{\mathbf{C}}(t,t') + \frac{1}{T}(\mathbf{1}+\boldsymbol{\Gamma})\widetilde{\mathbf{C}}_p(t,t')(1-q) \\
&\quad + \int_0^t \mathrm{d}s \widetilde{\boldsymbol{\Sigma}}_C(t,s)\widetilde{\mathbf{R}}(s,t') + \int_0^t \mathrm{d}s \widetilde{\boldsymbol{\Sigma}}_R(t,s)\widetilde{\mathbf{C}}(s,t')
\end{aligned} \tag{26}$$

(25), (24) and (26) describe the asymptotic dynamics of the $p$-spin under the Ichiki-Ohzeki dynamics.

We now proceed to implement the *ansatz* described in (18) for the slow contribution to $\mathbf{C}$ and $\mathbf{R}$. The long time value of the Lagrange multiplier reads

$$\boldsymbol{\mu}_\infty = \mathbf{1}\left[T + \frac{p^2}{2}\int_0^t \mathrm{d}s \widetilde{C}^{p-1}(t,s)\widetilde{R}(t,s) + \frac{p}{2T}(1-q^p)\right]. \tag{27}$$

We can relate the value of $\mu_\infty$ to the plateau value $q\mathbf{1} = \lim_{\tau\to+\infty} C_f(\tau)$ by taking the long time limit and the diagonal part of (25), thus obtaining $\boldsymbol{\mu}_\infty = \mu_\infty \mathbf{1}$ where

$$\mu_\infty = \frac{p}{2T}(1-q^{p-1}) + \frac{T}{1-q}. \tag{28}$$

This equation does not depend on $\gamma$, and is the same as the one obtained for $\gamma = 0$ (Eq. (7) of [11]).

Substitution of (18) and (28) in (26) finally yields

$$\begin{aligned} 0 &= \left[-\frac{T}{1-q} + \frac{p(p-1)}{2T}(1-q)\widetilde{C}(t,t')^{p-2}\right]\widetilde{R}(t,t') + \frac{p(p-1)}{2}\int_{t'}^t \mathrm{d}s \widetilde{C}(t,s)^{p-2}\widetilde{R}(t,s)\widetilde{R}(s,t') \\ 0 &= \left[-\frac{T}{1-q} + \frac{p}{2T}(1-q)\widetilde{C}(t,t')^{p-2}\right]\widetilde{C}(t,t') \\ &+ \frac{p}{2}\int_{t'}^t \mathrm{d}s \widetilde{C}(t,s)^{p-1}\widetilde{R}(s,t') + \frac{p(p-1)}{2}\int_0^t \mathrm{d}s \widetilde{C}(t,s)^{p-2}\widetilde{R}(t,s)\widetilde{C}(s,t'). \end{aligned} \tag{29}$$

Again, these equations are the same as the one obtained for $\gamma = 0$ (see Eqs. (22-23) of [12]). This establishes time-reparametrization invariance for the driven $p$-spin *à la* Ichiki-Ozehki.